\documentstyle[aas2pp4]{article}

\lefthead{Shimasaku & Fukugita}
\righthead{Galaxy Evolution}

\begin{document}


\newcommand\vs{\vspace{10pt}}
\newcommand\hs{\hspace{5pt}}
\newcommand\ite{\par\hang\textindent}
\newcommand\iteite{\par\indent \hangindent2\parindent \textindent}
\newcommand\lsim{\ \raise.46ex\hbox{$<$} 
 \kern-9.325pt \lower.47ex\hbox{$\sim$}\ }
\newcommand\gsim{\ \raise.46ex\hbox{$>$} 
 \kern-9.325pt \lower.47ex\hbox{$\sim$}\ }

\newcommand\astroph{astro-ph}

\def\BJ{B_{\rm J}}
\def\I814{I_{\rm F814W}}
\def\Mgas{M_{\rm gas}}
\def\MHI{M_{\rm HI}}
\def\Mb{M_{\rm baryon}}
\def\Ms{M_{\rm star}}
\def\zmed{z_{\rm med}}
 
\title{The History of Galaxies and Galaxy Number Counts}

\author{K. Shimasaku$^{1,2}$ and M. Fukugita$^{3,4}$}
\affil{$^1$ Department of Astronomy, University of Tokyo, Tokyo 113, 
Japan}
\affil{$^2$ Research Center for the Early Universe, 
       University of Tokyo, Tokyo 113, Japan}
\affil{$^3$ Institute for Advanced Study, Princeton, NJ 08540}
\affil{$^4$ Institute for Cosmic Ray Research, University of Tokyo, 
3-2-1, Midoricho, Tanashi, Tokyo 188, Japan}

%
%
 
\begin{abstract}
A simple quantitative model is presented for the history of galaxies to
explain galaxy number counts, redshift distributions and some other
related observations. We first infer that irregular
galaxies and the disks of spiral galaxies are young, probably
formed at $z\approx 0.5-2$ from a simultaneous
consideration of colours and gas content
under a moderate assumption on the star formation history.
Assuming that elliptical galaxies and bulges of spiral galaxies,
both  called spheroids in the discussion, had formed early in the
universe, the resulting scenario is that spiral galaxies 
formed as intergalactic gas accreting onto pre-existing bulges 
mostly at $z\approx 1-2$; irregular galaxies as seen today formed
by aggregation of clouds at $z\approx 0.5-1.5$.
Taking the formation epochs thus estimated into account,
we construct a model for the history of galaxies employing
a stellar population synthesis model.
We assume that the number of galaxies
does not change except that some of them (irregulars) were newly born,
and use a morphology-dependent local luminosity function to constrain
the number of galaxies. We represent the galaxies by E/S0, Sab, 
Sc and Irr; low luminosity dwarfs or any objects unobservable today 
do not play a role in our considerations.
In our model, spheroids follow passive evolution and the luminosity of
spiral galaxies evolves only very slowly for a wide redshift
interval due to a counterbalance between fading stars and new star 
formation from the gas replenished from intergalactic space.
Irregular galaxies evolve moderately fast for $z<1$.
The predictions of the model are compared with the observation of galaxy
number counts and redshift distributions
for the $B$, $I$ and $K$ colour bands. We show that $K$ band
observations are largely controlled by spheroids, 
which make them particularly suitable to study cosmology. 
We argue that $\Omega=1$ models are disfavoured, unless
the basic assumptions of the present model are abandoned.
The $K$ band observations reach quite
high redshift: for instance observations at $K$=23 mag may
explore the formation epoch, which could be as high as $z>5$.
On the other hand, galaxies observed in the $B$
band are dominated  by disks and
irregulars, spheroids making a very small contribution.
It is shown that young irregular galaxies cause the  steep slope of the 
counts. The fraction of irregular galaxies
increases with decreasing brightness: at $B=24$ mag, they
contribute as much as spiral galaxies.
Thus, ``the faint blue galaxy problem'' is solved by invoking
young galaxies.
This interpretation is corroborated by a comparison of
our prediction with the morphologically-classified galaxy counts in the
$I$ band. We do not invoke sporadic star bursting:
star formation
takes place steadily as does today, but galaxies (especially irregulars)
are gaseous at higher redshift, and hence star formation
is much more active than today.
Consistency is also shown with the constraint on the luminosity
evolution from a Mg II quasar-absorption-line selected sample.
We estimate that
2/3 of the baryons in stars are stored in spheroids and
1/3 in disks, only $<10$\% being in irregular galaxies.
The amount of baryons in disk stars is increasing, as they form
to $\Omega_b\sim0.001$, which just offsets the decrease of neutral
gas towards the present epoch, as inferred from
quasar absorption line surveys.
\end{abstract}
 
 
%
%

\section{Introduction}
 
When galaxies formed and how they  evolved
are among the central issues of cosmology today.
The traditional clue to this problem has been provided by
the number count of
galaxies as a function of apparent magnitude.
Early observations in the blue band showed  that the
number of galaxies per unit angular area
increases faster than is expected in simple models
for fainter magnitude
(Koo \& Kron 1992, for a review).
It has become clear, however, that the number count
alone does not allow an unambiguous interpretation.
Over the last decade, much effort has been expended to
obtain redshift of faint galaxies (Broadhurst, Ellis \&
Shanks 1988, hereafter BES;
Colless et al. 1990; 1993; Lilly, Cowie \& Gardner 1991;
Koo \& Kron 1992; Songaila et al.
1994; Glazebrook et al. 1995a; Lilly et al. 1995; Ellis et al. 1996).
These redshift surveys have revealed, when combined with
the count data, that the distribution of galaxies in the universe is
much more puzzling than it looked:
the excessive number of galaxies observed in the number count 
seemed to be most simply explained by assuming luminosity
evolution of galaxies (e.g., Koo \& Kron 1992), but
the redshift surveys have indicated that the shape of
the redshift distribution appears almost
as is predicted with a no-evolution model, with the normalization,
however, being larger by a factor of two. Simple models
with all galaxies undergoing substantial luminosity evolution,
which account for the steep slope of
the number count, predict too many galaxies having high redshift
to be consistent with the observations  (e.g., Ellis 1990;
Lilly 1993) (though this seems to have been somewhat overstated).
 
There have been a number of attempts to solve this problem:
some authors speculated that the excess count is due to a new
population of galaxies which have undergone
star burst at rather low redshift (BES; Cowie, Songaila and Hu 1991). 
Babul \& Rees (1992) and Babul \& Ferguson (1996) interpreted
this new population as  dwarf galaxies from collapsed Lyman $\alpha$ 
clouds, eventually faded into galaxies making up an invisible
population.
Some other authors have assumed a heuristic morphology-dependent
luminosity evolution so that subluminous galaxies evolve fast
to explain high counts while keeping the shape of redshift
distribution as is in the no-evolution model, rather than introducing
a new population  (BES; Lilly 1993;
Phillips \& Driver 1995).
Another class of explanation invokes mergers of
galaxies with a very high
merging rate (Guiderdoni \& Rocca-Volmerange 1991; Broadhurst,
Ellis \& Glazebrook 1992; Carlberg 1992).
The required rate to explain the steep slope of the $B$ band
count is so high that a typical galaxy gains 50\% of mass over
the past 5 Gyr.
More recently, Gronwall \& Koo
(1995) suggested that the presence of an abundant non-evolving very blue
dwarf population accounts for both counts and redshift data,
although the meaning of this blue population is not clear to us.
 
Observational data have now accumulated for both counts and redshifts
over a wide range of colours from $B$ to near infrared $K$ band, and
the deepest redshift reaches beyond $z=1$ (Glazebrook et al. 1995a;
Songaila et al. 1994; Lilly et al. 1995).
A large number of redshifts contained in the survey have enabled
a direct construction of the $B$ band luminosity function as a function
of redshift (Lilly et al. 1995; Ellis et al. 1996; see also Eales 1993).
This prompts us to think that we already have enough data 
to understand the evolution of galaxies below $z \sim 1$.
In addition, more varieties of clues have become available
to understand the nature of faint galaxies.
Sharp images of galaxies obtained with the HST have enabled a study of
morphology to $I$=24 mag, giving us a direct probe concerning
how the faint galaxies look like (Driver, Windhorst \& Griffiths 1995;
Glazebrook et al. 1995b; Abraham et al. 1996; Driver et al. 1996). 
The work has shown that the fraction of irregular galaxies
increases sharply with magnitude beyond $I>18-20$ mag. The same
observations indicate that giant galaxies evolve only slowly,
although they do indicate some evolution in colour, brightness and
the detail of morphology.
Another piece of evidence that giant
galaxies evolve only slowly, if at all, comes from a
Mg II quasar-absorption-line selected sample
(Steidel, Dickinson \& Persson 1994).  The luminosity of
galaxies that yield Mg II absorption lines for a quasar
in their vicinity is remarkably constant both in the
rest-frame $B$ and $K$ bands from $z=0.3$ to $\approx$1.
 
There are also a number of observations that tell us about
the content of galaxies.  The information that would directly constrain
the model for evolution of galaxies is the estimate of the
global star formation rate at $z=0$ (Gallego et al. 1995)
and high $z$ (Cowie, Hu \& Songaila 1995; 
see also Madau et al. 1996 for a more recent work) 
from the strength of H$\alpha$ or [O II]$\lambda 3727$ emission lines. 
This quantifies the
inference from early observations concerning the increase of
star formation activity in the past (BES).  Yet another useful
information is given by the measurement of the neutral hydrogen
abundance in the universe as a function of redshift (Lanzetta,
Wolfe \& Turnshek 1995; Storrie-Lombardi, McMahon \& Irwin 1996).  
The data show clear depletion of HI gas with time, suggesting that
this gas has been used for star formation.
This information can be used to
constrain the total volume emissivity of galaxies
(see, Pei \& Fall 1995), on which direct information is also
available from galaxy observations (Lilly et al. 1996).
 
From accumulating observations we now have a reasonable picture 
for the evolution of galaxies (Fukugita, Hogan \& Peebles 1996):
giant galaxies were already mature at $z\sim 1$ and evolve only
slowly lower than  this redshift, whereas subluminous galaxies evolve
rapidly in the same redshift range.
The latter are probably young galaxies formed close to $z\sim 1$.
It has also been speculated that most of elliptical galaxies
and bulges of the spiral galaxies, which altogether we call spheroids,
formed before $z\gsim 3$, and spiral disks were assembled in
the redshift range from $z\sim 3$ to 1.
 
In this paper we present a simple but quantitative model for
the history of galaxies, which is broadly consistent with
the observations. The model employs 
a population synthesis model of stars. 
In the traditional attempts, galaxies are assumed to have fully 
assembled at an early epoch of the universe (Searle,
Sargent \& Bagnuolo 1973; Sandage 1986; Yoshii \& Takahara 1988),
and the star formation rates are adjusted to reproduce galaxy colours
today.  In our attempt it is crucial to properly build in the epoch
of the formation of galaxies in the stellar population synthesis
model.  For this purpose we first show that one can estimate the
age of the galaxies
as seen today from a simultaneous consideration of
the colour and gas content of galaxies
with the aid of moderate assumptions on the star
formation history, and infer the age of irregular galaxies and
also of disks of spiral galaxies. 
For elliptical galaxies, and also bulges of spiral galaxies from
family resemblance (both are 
called spheroids), we adopt the conventional view that they
formed at high $z$ as argued by Eggen, Lynden-Bell
and Sandage (1962). 
We assume that the number of galaxies is conserved since they were
born, and it is constrained by  the local type-dependent
luminosity function. 
We calculate the evolution of bulges and disks
separately and assume delayed formation of disks and irregular
galaxies. The spheroids are supposed to follow 
the passive evolution to now.
Galaxies are grouped into four morphology
types: E/S0, Sab, Sc, and Sdm-Irr which is simply referred
to as Irr in this paper; low luminosity dwarf galaxies do not play 
an essential role in our argument. 
The mix of morphological types of galaxies is not constant, but 
depends on absolute luminosity.
 
The predictions of this model are compared with the observation
of the galaxy number count and the redshift distribution in $B$, $I$
and $K$ bands. We do not discuss the $U$ band count, since we
cannot model reliable $K$- and $E$-corrections for this band,
and $U$ band data are too sensitive to the detail of galaxy activity.
We discuss the evolution of gas and star content of galaxies
predicted in the model. Through these comparisons with the observations
we may conclude that the present model is reasonably well constrained
for $z\lsim 1-1.5$. The prediction for $z>1.5$, however, is
basically an extrapolation, and ill-constrained.
A number of checkpoints are also discussed to prove or disprove the
validity of the model and, more importantly, 
underlying assumptions taken in this paper.
 
We do not consider the no-evolution model seriously,
since it is clearly unphysical. We know that elliptical galaxies
{\it must} evolve even in the absence of new star formation after
the burst, since the stars in the main sequence branch
continuously evolve to giants, and the total luminosity should
decrease with time (Tinsley 1972). 
This basic feature is a solid prediction of
the stellar population synthesis model, although quantitative details
depend on models, in particular on assumed initial mass functions.
For spiral galaxies, there are more uncertainties:
our model corresponds to the case where luminosity changes
very little with time: the decrease
of light due to evolution of stars is compensated by
newly born stars formed from the gas replenished into the disk.

We note that the present model is quite different from ``CDM
cosmologies'', 
where morphologies of galaxies change
from time to time due to mergers down to low redshift
(Kauffmann, Guiderdoni \& White 1994; Cole et al. 1994).
That is, elliptical galaxies
are not quite old, but formed by collision of, for instance, two
spiral galaxies at rather low redshift (Toomre 1977).
 
One of the original reasons for interest in faint galaxy counts is
that it may offer a chance to test world geometry (Sandage 1961).
A number of attempts have been made aiming at this goal (e.g.,
Yoshii \& Takahara 1988; Fukugita et al. 1990; 
Gardner, Cowie \& Wainscoat 1993; Gronwall \& Koo 1995). 
The reliability of the results, however, is
strongly affected by the uncertainties of the model.
In this paper we discuss
where the predictions are reasonably robust.
In particular we discuss the possibility to test the
presence or absence of the cosmological constant.
 
In the next section we discuss how to estimate the ages of galaxies from
their colours and gas content. The details of the model and
input assumptions are discussed in section 3.  In section 4 we present
the predictions of the model and compare them with the observations in
$K$, $B$ and $I$ bands, and with the constraint on evolution 
from a Mg II quasar absorption line selected sample. 
The evolution of baryons in stars and in gas is also discussed. 
Section 5 is devoted to discussion and summary of the problems.
 
%
%

\begin{figure}[t]
\plotone{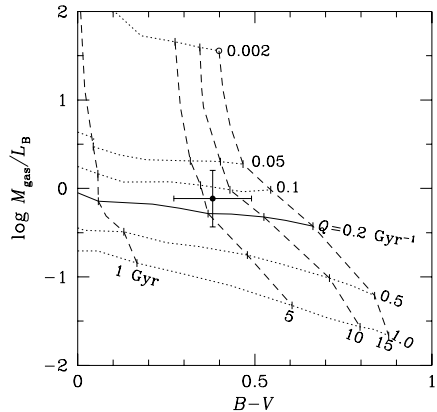}
\caption
{Irregular galaxies in the plane $B-V$ vs the
gas content normalized by $B$ luminosity $M_{\rm gas}/L_B$. 
Error bars show the variance of the distribution. 
Data are taken from Buta et al. (1994). Solid
and dotted curves show evolutionary tracks with the
star formation rate specified by $Q$ ($Q=0.2$ is the value at
the solar neighbourhood). The terminating point with the
conventional model (``old irregular galaxies'') is indicated by the 
open circle. 
The dashed lines indicate isochronal contours, 1, 5, 10 and 15 Gyr
from the birth of galaxies.
\label{fig1}}
\end{figure} 

\section{Age and star formation rate of galaxies}
\subsection{Irregular galaxies}
 
The two fundamental parameters that control the evolution of the
stellar population are the age of the galaxy $t_G$ 
and the star formation rate $Q$ (Gyr$^{-1}$), 
provided that the initial mass function is fixed. 
We can obtain useful information
concerning these parameters by plotting galaxies in the two
dimensional plane, integrated galaxy colour versus gas. 
In Fig. 1 we show
a diagram for irregular galaxies plotted in $B-V$ versus
$M_{\rm gas}/L_B$, where $M_{\rm gas}$ is mass of the neutral atomic
gas estimated from the
integrated 21cm HI flux, corrected for helium abundance of 25\% in mass,
with the aid of the formula given in the
{\it Third Reference Catalogue of Bright Galaxies} (de Vaucouleurs
et al. 1991, hereafter RC3). Molecular hydrogen is ignored.
The data are taken from Buta et al.
(1994), but a very similar figure can be obtained using the data
given in RC3.  The bars indicate one standard
deviation of the distribution.
 
Evolutionary tracks are presented in the figure for a given
$Q$ using a conventional evolution model of galaxies, where
star formation is assumed to take place at a constant rate against
gas mass.
For the present calculation,
we adopt the closed model of Arimoto \& Yoshii
(1986), but other models also yield
similar results. The track represented by a solid line
corresponds to $Q=0.2$ Gyr$^{-1}$, which is the star
formation rate in the solar neighbourhood, and is also consistent
with the estimate for Sb-Sbc galaxies (Roberts 1963; Smith,
Biermann \& Mezger 1978; Kennicutt 1983; Sandage 1986).
The ticks on the track indicate the age
of galaxies, 1, 5, 10 and 15 Gyr after star formation began,
so that dashed curves represent isochronal contours.
We note that the traditional model, where irregular galaxies
formed as old as the universe, assumes a very low
$Q$ ($Q\simeq0.002$ Gyr$^{-1}$, say) 
in order to obtain their blue colour ($B-V\simeq0.4$).
This terminating point of the $Q=0.002$ track
is indicated by an open circle.
The family of evolutionary tracks indicates that one can
obtain the same colour with a larger $Q$ if the age of galaxies
is younger; the gas fraction determines the age.
The data point for irregular galaxies shows that they are
as young as $t_G=$3--10 Gyr and $Q\approx 0.1-0.2$ Gyr$^{-1}$,
which is close to the solar neighbourhood value.
The prediction of the traditional model, as presented by
the open circle, is clearly inconsistent with the observation.
We emphasize the large ``error bars'', which indicate that
the irregular galaxies observed today have not had a
coeval formation at all, with an age spread of well over 4 to 12 Gyr.
 
One attractive feature with a larger $Q$, which is consistent with the
value for spiral galaxies, is that it supports the validity of the
Schmidt law, which states that the star formation rate is
proportional
to the HI gas amount to some power (Schmidt 1959).
The approximate validity of the Schmidt law with power law index
close to unity is supported
by an analysis of emission line features (Kennicutt 1989).
The traditional value
$Q=0.002$ Gyr$^{-1}$ requires an unusually strong
suppression of star formation in irregular galaxies,
the surface brightness of which is not necessarily very low.
 
We do not consider intermittent star bursts, which are often
invoked in the literature to account for faint blue galaxies.
We consider that the star formation activity is steady and continuous
at least for the majority of galaxies.
The galaxies are highly gaseous shortly after their birth, 
and the star formation rate {\it per galaxy} can be 4 times than it
is today.  Those galaxies are seen as if they are undergoing
star burst activity.
 
We conclude that irregular galaxies observed today
formed at around $z\sim 1$, although some irregulars are
formed as low as $z\sim 0.5$ or less, and some others as high
as $z\sim 2$.
We do not mean, however, that the irregular galaxies do not 
form at high $z$.
Most of the galaxies formed at higher $z$,
might already have been faded or accreted onto giant galaxies
and they do not show up in our local sample of irregulars.
 
%
%

\begin{figure}[t]
\plotone{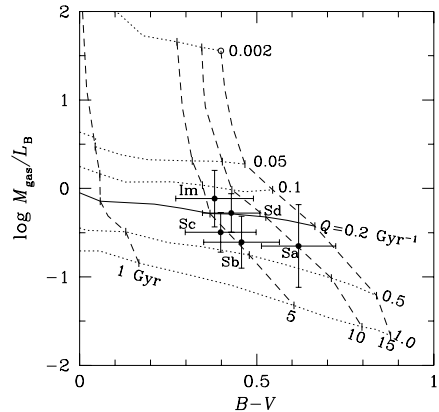}
\caption
{Same as Fig. 1, but for the disk components of
Sa, Sb, Sc galaxies. The data for Sd and Im galaxies are also
added.  Data are taken from RC3.
\label{fig2}}
\end{figure}
 
\subsection{Disks of spiral galaxies}
 
A similar analysis can be applied to disk components of spiral
galaxies. Figure 2 shows $B-V$ versus $\Mgas/L_B$ plot for
spiral disks, where both $B-V$ and $L_B$ are corrected for
the light from bulge component by assuming average
disk/bulge ratios that depend on morphology (see section 3.1 below).
The data are taken from RC3.
A plot is also made for Irr and Sd for comparison.
The tracks are calculated with a closed model.
 
This figure shows that the disks of late-type
spiral galaxies are also young with ages around $\approx$5-12 Gyr
allowing for a large scatter, if a slight upward shift of the data 
due to the neglect of H$_2$ gas is taken into account.
The plot for Sa indicates that some of these might be as
old as the spheroids.
 
The age of the disk of Sb - Sc galaxies inferred this way is consistent
with the age of the oldest disk stars (7.5--11 Gyr) estimated
from nearby white dwarfs (Winget et al. 1987; Wood 1992).
 
For bulges of spiral galaxies it is reasonable to
suppose that they formed very early, $z \gsim 3$, from family
resemblance with elliptical galaxies, if they indeed
formed so. The resulting picture is that spiral disks
formed as intergalactic gas accreted onto preexisting bulges and
star formation began in the disk at $z\sim 1-3$.
 
%
%

\section{The model}
 
\subsection{Galaxy formation epoch and the evolution model}
 
We assume that spheroids (ellipticals and bulges of spiral galaxies)
formed at high $z$ ($z>5$) by a single
short burst event. The stars have to be formed from the initially
collapsing gas faster than the collapse time to avoid disk
formation (Eggen et al. 1962).
The duration of the burst is taken to be  0.125 Gyr,
i.e., the star formation rate $Q_*$ (hereinafter the star
formation rate is denoted as $Q_*$) being effectively
40 times the solar neighbourhood value (e.g., Sandage 1986).
These galaxies
evolve passively after the burst. The problem inherent in the
presently available
stellar population synthesis models is that they do not reproduce
the UV spectrum of average elliptical galaxies today. It shows a
significant UV component and frequently an upturn shortward of
2000\AA~ (Bertola, Capaccioli \& Oke 1982; Burstein et al. 1988). 
In the models of passive evolution the UV component rapidly dies off
a few Gyr after the burst.  The UV component of elliptical galaxies,
however, plays very little role in the quantities that we discuss
in the present paper. No effect is expected in the $K$ band
until one samples galaxies above $z \sim 6$. In this pass band the
effect is visible only in the high redshift tail of the redshift
distribution at $K>23$ mag.  For the $I$ band the effect appears
only above $z>2$, and a minor change is expected only for $I\gsim 25$.
For the $B$ band we expect some effect for redshifts is as
low as 0.5, but the spheroids play only a minor role in this
pass band, and the presence or absence of the UV component
does not change any predictions, unless elliptical
galaxies are specifically
selected from the sample.  For this reason we can ignore the
problem of the disagreement of the passive evolution model with
the UV observations
\footnote{
This component becomes important when one discusses a small
high redshift tail of the $z$ distribution in the optical
pass bands. A typical example is the ``Lyman continuum break
galaxies'' recently discovered by Steidel and collaborators (1996) in
a sample selected in the $R$ band. Such observations just sample those
galaxies that are dominantly controlled by the UV components of
spheroids.  
}.

We represent spiral galaxies with two morphological types, Sab and Sc.
We assume that they consist of old spheroids and disks which formed
typically at $z\sim1-2$. We adopt
for the disk components a model where gas continuously
accretes onto spheroids, making spiral disks with $e$-folding
time $Q_{\rm in}^{-1}=5$ Gyr ($Q_{\rm in}^{-1}$ is the
inverse of the infall rate),
which corresponds to $z\sim1-2$. The star formation
rate $Q_*$ is set equal to the
value in the solar neighbourhood, $Q_*=0.2$ Gyr$^{-1}$,
irrespective of galaxy types.
We use the stellar population synthesis model of Arimoto, Yoshii \&
Takahara (1992) to accommodate gas infall.
This infall model gives evolutionary tracks significantly different
from the ones based on a closed model shown in Fig. 2 
for a high $Q_*$ case, but the
self-consistency between the input and output is maintained for
(age, star formation rate) versus (colour, age).
The bulge
fraction in the blue light is taken to be 0.3 for Sab and
0.1 for Sc in agreement with the analysis of Kent (1985)
after the colour transformation.
We remark that we adopt here a model of disk formation
exponential in time with $e$-folding time $Q_{\rm in}$ for simplicity.
The formation epoch of irregular galaxies is taken to be 8 Gyr
back from the present for simplicity of the model, 
although the actual formation epoch spreads over a wide range of age.
We assume a universal star formation rate,
the same value as that for disks.
 
We take the age of the universe to be 15 Gyr, but only minor
modifications are caused if we take it to be 13.5 Gyr (e.g.,
E galaxies are slightly bluer, and evolution is a little stronger etc.).
The Hubble constant enters into our scenario only through the
evolution time-scale of stars. We assume the Hubble constant to
have the value specified by the assumed age of the universe and the 
specific cosmology we consider.

\begin{table}
\caption{Assumed history of star formation in galaxy.}
\begin{tabular}{cccc}
\hline
\hline
      & $t_G$ (Gyr)  &  $Q_*^{-1}$ (Gyr) & $Q_{\rm in}^{-1}$ (Gyr) \\
\hline
spheroids  & 15 & 0.125 (burst) &  instantaneous \\
disks      & 15 & 5             &  5             \\
irregulars & 8  & 5             &  instantaneous \\
\hline
\end{tabular}
\label{tab1}
\end{table}

\vspace{10pt}
 
\begin{table}
\caption{Properties of galaxies at the present epoch.}
\begin{tabular}{crrrrr}
\hline
\hline
                       & E/S0  &  Sab  &  Sc   &  Im   &  disk \\
\hline
bulge fraction ($L_B$) &  1    &  0.3  &  0.1  &  0    &  ---  \\
bulge fraction ($L_K$) &  1    &  0.5  &  0.15 &  0    &  ---  \\
$\Mb/L_B$              &  8.4  &  4.7  &  3.6  &  1.8  &  3.1  \\
gas fraction           &  0    &  0.10 & 0.18  &  0.32 & 0.22  \\
$B-V$                  & 0.98  &  0.68 & 0.57  & 0.46  & 0.51  \\
$V-R$                  & 0.90  &  0.73 & 0.68  & 0.61  & 0.65  \\
$V-I$                  & 1.69  &  1.42 & 1.32  & 1.18  & 1.27  \\
$V-K$                  & 3.11  &  2.87 & 2.66  & 2.42  & 2.56  \\
\hline
\end{tabular}
\label{tab2}
\end{table}
 
The parameters of our model are summarized in Table 1. Some
basic predictions for galaxy properties, bulge fraction
(input), $\Mb/L_B$, gas fraction, $B-V$, $V-R$, $V-I$ and $V-K$ colours
at $z=0$, are presented in Table 2 (discussion is deferred to
later sections). We use the standard Johnson-Morgan-Cousins colour band
system, unless otherwise explicitly denoted. The evolution
of the light emitted by E, Sc and Irr galaxies is shown in Fig. 3
for the rest frame $B$ and $K$ bands.  Spheroids follow the standard
passive evolution for both pass bands.  The brightening of
0.3 mag in $B$ between $z=0.3$ and $0.6$ is consistent with the 
observation for red galaxies in the Lilly et al. (1995) sample, 
in particular for $q_0=0$ cosmology.
Sc galaxies hardly
evolve in the $K$ band. In the $B$ band, weak evolution is visible only
between $z=0$ and 0.5, only in the epoch after gas infall has ceased.
The evolution between $z=0$ and 0.5 is about a half mag.
(This is compared to Lilly et al. 1995, who reported 1 mag
evolution of $L_B^*$ from $z=0$ to $z=0.6$; they have not seen
a change of $L^*$ above $z>0.75$, which is consistent with our model).

\begin{figure}
\plotone{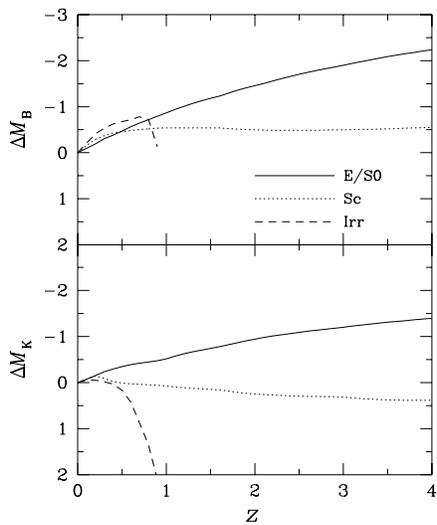}
\caption
{Evolution of the rest frame $B$ and $K$ band 
luminosity for E/S0, Sc and Irr galaxies in our model.
\label{fig3}}
\end{figure}
 
Below $z\le 1$, the galaxies that have undergone 
the fastest evolution are Irr, 
but the total increment in the $B$ band luminosity 
is only 0.8 mag between $z=0$ and 0.8.
Of course, the evolution is much faster in the UV light,
and the $E$ correction
for the $B$ count amounts to $\approx$2 mag in the same redshift range.
This is almost the maximum evolution that can be attained 
with a conventional stellar population synthesis model.
Lilly (1993) and Phillips \& Driver (1995) assumed
significantly faster evolution ($\Delta m=4$ mag between $z=0$ and 0.4,
and 2.5 mag between $z=0$ and 0.5, respectively).
 
%
%

\subsection{Local luminosity function}
 
Another input to the model that deserves discussion 
is the local luminosity function (LF). 
We need the morphology-type dependent luminosity
function.  The only case where we know such
LF reasonably well is that for the Virgo cluster (Bingelli, Sandage
\& Tammann 1988).
The LFs, when galaxies are classified into morphological types,
show cutoffs in
both bright and faint ends.
On the other hand, we know reasonably well that the LF summed over
all morphological types behaves like the Schechter function
(Loveday et al. 1992; Lin et al. 1996, Ellis et al. 1996), although
its faint end is uncertain.
We adopt as a fiducial LF the one
reconstructed by Lilly (1993), which takes account of all these features
(reproduced in Fig. 4). In this figure Sdm and dIrr are plotted
separately, but our results do not depend on whether we keep 
the dIrr component or drop it.  
Hereinafter, we simply use terminology of Irr for
the sum of Sdm and dIrr.
The total LF is represented by a Schechter function
with $\alpha=-1.15$, $M_{\rm B}^*=-19.6+5\log h$ mag, 
and $\phi^*=0.018h^3$ (Mpc)$^{-3}$, 
where an upward shift by 0.1 dex is made for the
normalization.  This corresponds to local luminosity density ${\cal L}=
2.11\times 10^8\ h L_\odot$(Mpc)$^{-3}$.
We remark that the normalization of the LF has a substantial
uncertainty.
Our adopted normalization  is consistent with those of
Efstathiou, Ellis \& Peterson (1988) and of Ellis et al. (1996) with
the luminosity density 1.93 and 2.05$\times 10^8hL_\odot$(Mpc)$^{-3}$,
respectively, but is higher than that of Loveday et al. (1992),
$1.35\times 10^8\ h L_\odot$(Mpc)$^{-3}$. We see a 50\% uncertainty
in ${\cal L}$ among the widely accepted LFs 
\footnote{This
is smaller than the uncertainty seen in $\phi^*$. The determination
of  $\phi^*$ is correlated with $L^*$, and the resulting 
uncertainty in  ${\cal L}$ is smaller than that in  $\phi^*$}.
As we discuss later the interpretation of the number count
is affected by this uncertainty.

\begin{figure}[t]
\plotone{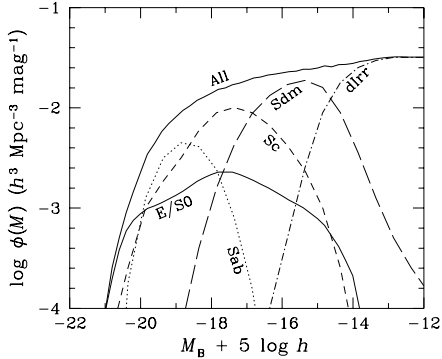}
\caption
{Luminosity function used in the present paper.
It is adopted from Lilly (1993), but the normalization is 
shifted upward by 0.1 dex. 
\label{fig4}}
\end{figure}
 
We remark that the slope of the faint end is not
very important to the overall shape of the number counts,
unless the faint tail of the LF rises so sharply that it leads to
a substantial enhancement of the integral, i.e., $\alpha\lsim -1.6$.
The contribution to the number count (see eq. [1] below)
is luminosity weighted, as $\sim L^{*3/2}\phi_*$, 
so that the behaviour of the number count is basically
determined by galaxies with the characteristic luminosity.

On the other hand, the LF of irregular galaxies may
well be higher by a factor of two or more, which increases the
contribution from irregular galaxies by this amount. In this paper
we adopt the above LF as a fiducial choice and then discuss whether
the modifications necessary to explain the observation, if any, are
within the uncertainties of the current LF,
rather than modifying the adopted LF.

%
%

\subsection{Cosmology}
 
The differential number count of galaxies is given by
$$n(m_\lambda,z)={\omega \over 4\pi}{dV \over dz}\phi(M_\lambda),
\eqno{(1)}$$
where $\omega$ is solid angle, $V$ is the comoving volume, $\phi$
is the LF, and the relation between
$m$ and $M$ in the $\lambda$-pass band is given by
$$m_\lambda=M_\lambda+K_\lambda+E_\lambda+5\log (d_L/10{\rm pc}),
\eqno{(2)}$$
with $d_L$ the luminosity distance.
The $K$ correction for the optical band is taken
from Fukugita, Shimasaku \& Ichikawa (1995).  We adopt the $K$
correction of Cowie et al. (1994) for the $K$ band.  There is
little difference among authors for $K$ corrections for any
colour bands other than far UV.
We also refer to Fukugita et al. (1995) for the transformation of
magnitudes between the different colour band systems, which is
extensively used in this paper.
The $E$ correction is calculated according to the  prescription
given in section 3.1 above
\footnote{For the spheroids, we assume
that the SED at wavelength shorter than $\lambda_U=$3000\AA~ did not
change in shape but scaled only in intensity with the
point at $\lambda_U$ to avoid too large an $E$ correction,
which is predicted by the pure passive evolution model
(Yoshii \& Takahara 1988). As discussed previously, however,
we only consider the predictions which are insensitive
to the UV SED of the spheroids.}.
The prediction is compared with 
the observation by integrating (1) over $dz$ and/or $dm$ for
appropriate ranges.
 
We consider a number of cosmological models when we
discuss the $K$ band count, which shows the largest power
in discriminating among cosmological models. For other colour bands 
we are confined mainly to two cases:
(i) open universe with $\Omega=0.1$, and
(ii) $\lambda$-dominated universe with $\Omega=0.1$ and $\lambda=0.9$,
as an extreme case, although such high $\lambda$ seems to be excluded
from the statistical lensing.
$\Omega=1$ models fail to reproduce the data for the $K$ band
count, and hence are omitted from most of our considerations.
 
%
%
 
\section{Predictions of the model}
 
\subsection{Galaxies in the $K$-band}
 
\noindent
{\it Number count}
 
Figure 5 shows galaxy number counts in the $K$ band. The data are
taken from Djorgovski et al. (1995), Gardner et al. (1993) and 
Soifer et al. (1994) for faint counts, 
and Glazebrook et al. (1994) and Mobasher, Ellis \& Sharples (1986) 
for brighter counts.
The curves are shown for four cosmology models: $\Omega=1$;
$\Omega=0.1$ open model; ($\Omega=0.1$, $\lambda=0.9$); and
($\Omega=0.3$, $\lambda=0.7$).
We can see gross agreement between the models and the observation
for open and $\lambda\ne0$ models. The predicted count for
the $\Omega=1$ model falls short of the observed count beyond 20 mag,
by a factor of 4 at 23 mag,
showing that $\Omega=1$ model is disfavoured. This conclusion would not
be modified unless there are unexpectedly many red dwarfs in the
local LF (so that $\alpha\approx -2$) at very faint magnitudes, or
the basic assumptions of the present model
(conserved number of galaxies; early spheroid formation etc.)
are abandoned; see the discussion section below.

\begin{figure}
\plotone{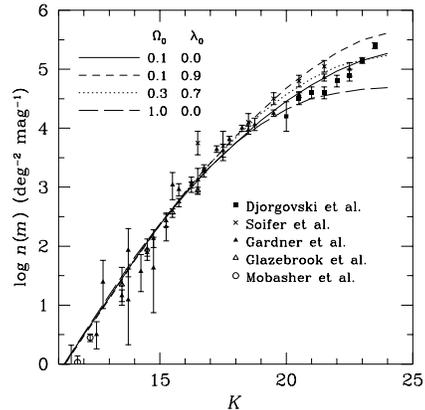}
\caption
{$K$ band galaxy number count, as compared with the
model predictions for four cosmology models.  Data are taken from
Djorgovski et al. (1995) (solid squares), Soifer et al. (1994)
(crosses), Gardner et al. (1993) (solid triangles), Glazebrook et al.
(1994) (open triangles) and Mobasher et al. (1986) (open circles).
\label{fig5}}
\end{figure}
 
Looking at details we note a few features: a good agreement at
bright magnitudes indicates that the normalization of the $K$ band
count is given correctly with the $B$ band LF and the colour
transformation without further adjustment.
The observed counts are slightly higher at $K=17-18$;
the data look as if there is a mild change in the slope 
at this magnitude, while such change is not accounted for by the
usual $K$ or $E$ correction.
Another feature is that the prediction of the 
$(\Omega=0.1, \lambda=0.9)$ model overshoots the data of 
Djorgovski et al. for the range 20-23 mag 
by a factor of $\sim 2$. 
On the other hand, this $\lambda$-dominated model just fits the 
data of Soifer et al.
The survey areas for these deep counts are only 1-2 square 
arcminutes for the above two surveys, and 16 square arcminutes 
in the Gardner et al. survey. 
It is likely that the data suffer from the effect of large-scale 
inhomogeneity beyond $\sqrt N$ error estimates.
For this reason we cannot yet rule out the $\lambda$-dominated
model from the analysis of these counts. A deep survey of larger area
is highly awaited in this regard.

We find that some small wiggle is induced to the $N(m)$ curve
if the epoch of burst-like spheroidal formation is as low as $z=5-6$.
The effect is more clearly visible in the redshift distribution. 
Nevertheless, the gross feature of $K$ band count is fairly 
insensitive to the model details; it is more sensitive to cosmology.
 
It is generally expected that the $K$ band light
is sensitive to the old population.
We show in Fig. 6 the fraction of light in the $K$ band arising
from spheroidal components for an $\Omega=0.1$ universe. (A similar
plot is also shown for the blue light for comparison.) This
figure shows that $\gsim60\%$ of light comes from spheroidal
components in the entire  magnitude range that concerns us,
confirming the expectation that $K$ band counts are a
good probe for old population.  Since we understand the evolution
of spheroids reasonably well, there is no much difference among
predictions of different authors. This makes $K$ counts
suitable to study cosmology, especially for a test for the cosmological
constant. The uncertainty present in the
data, however, does not allow us to make a conclusion.
For the blue light the fraction of spheroid contribution
is about 10\% for $B_J\gsim 22$ mag.

\begin{figure}
\plotone{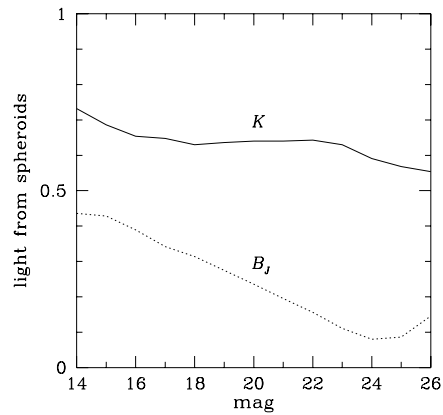}
\caption
{Fraction of light emitted by spheroids from galaxies
which are observed at specified magnitude. 
\label{fig6}}
\end{figure}
 
\smallskip
 
\noindent
{\it Redshift distribution}
 
Three panels in Fig. 7 show the redshift distribution for
the magnitude range (a) $K=17-18$, (b) $K=19-20$ and
(c) $K=23-23.5$.
The prediction is compared in (a)
with the $z$ survey data of Songaila et al. (1994) after 
normalizing them to the count data with the sample incompleteness
taken into account.
The figure shows good agreement for the global shape,
although the data are higher than the prediction by 20\%
(note that this is the magnitude where the disagreement between
prediction and data becomes maximum, as we noted above) and they
lack a high $z$ tail above $z\simeq 1$.
At this magnitude 15\% of galaxies are
not given redshift data, and it is conceivable that they are
missed by redshifting the characteristic [O II] line out of
the spectroscopy window, as it is more clearly seen in their
fainter magnitude sample, where basically no galaxies
above $z=1$ are catalogued. We remark that shoulders or wiggles
seen in summed redshift distribution are artefact of
representing galaxies with only four distinct types. Since morphology
is a continuous class,
these curves must be sufficiently smeared when we compare
them with the observation.

\begin{figure}
\plotone{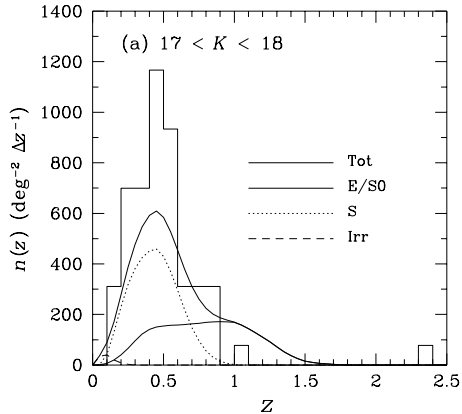}
\caption
{(a) Redshift distribution of galaxies selected with $K=17-18$ mag.
$\Delta z$ is taken to be 0.1. The prediction is for an open universe,
$\Omega=0.1$.
The data from Songaila et al. (1994), 
normalized to the count data, are also shown. 
Note that the completeness of this $z$ survey is 85\%.
\label{fig7}}
\end{figure}

\setcounter{figure}{6}
\begin{figure}
\plotone{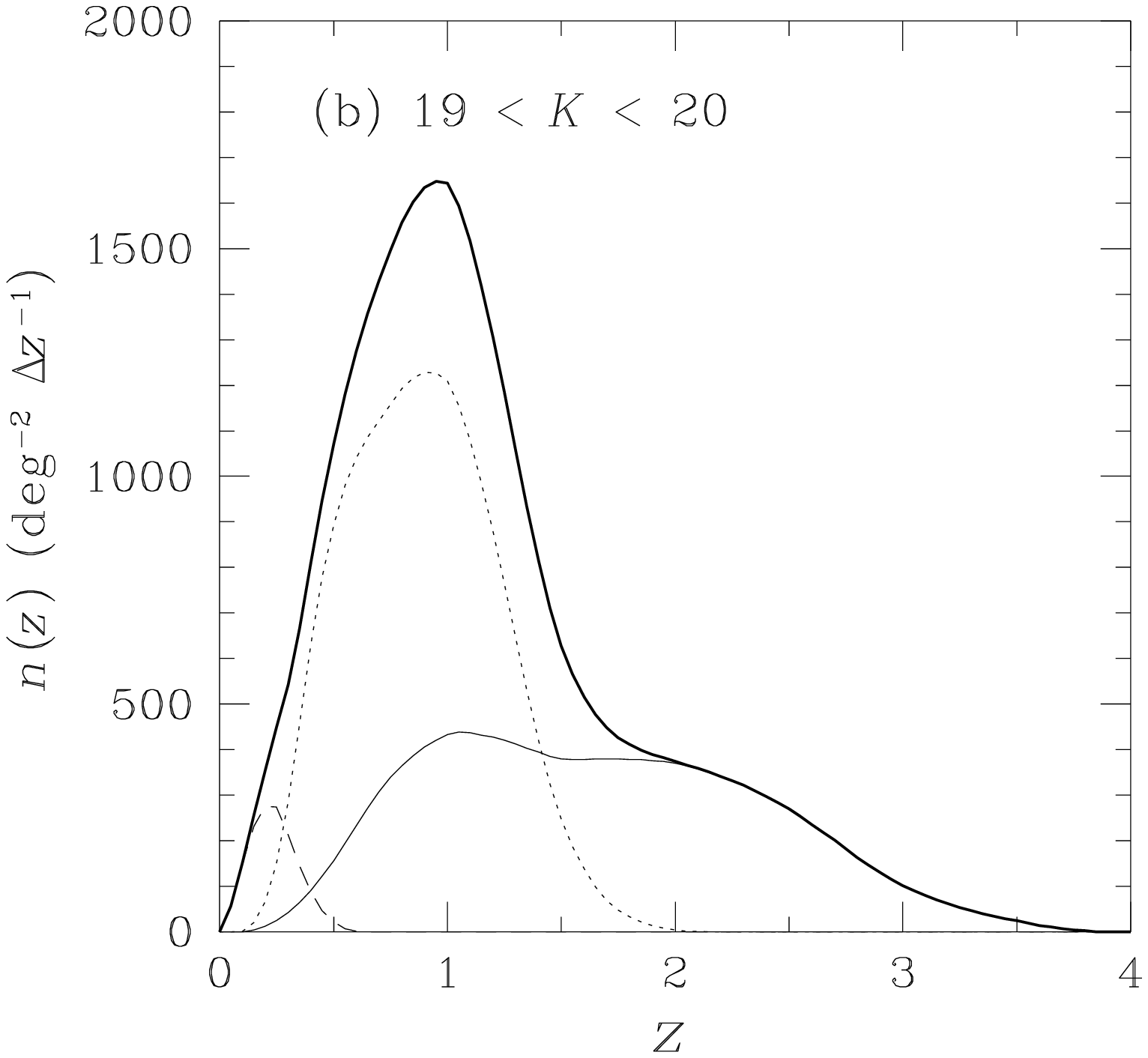}
\caption{(b) Same as (a), but for $K=19-20$ mag.}
\end{figure}

\setcounter{figure}{6}
\begin{figure}
\plotone{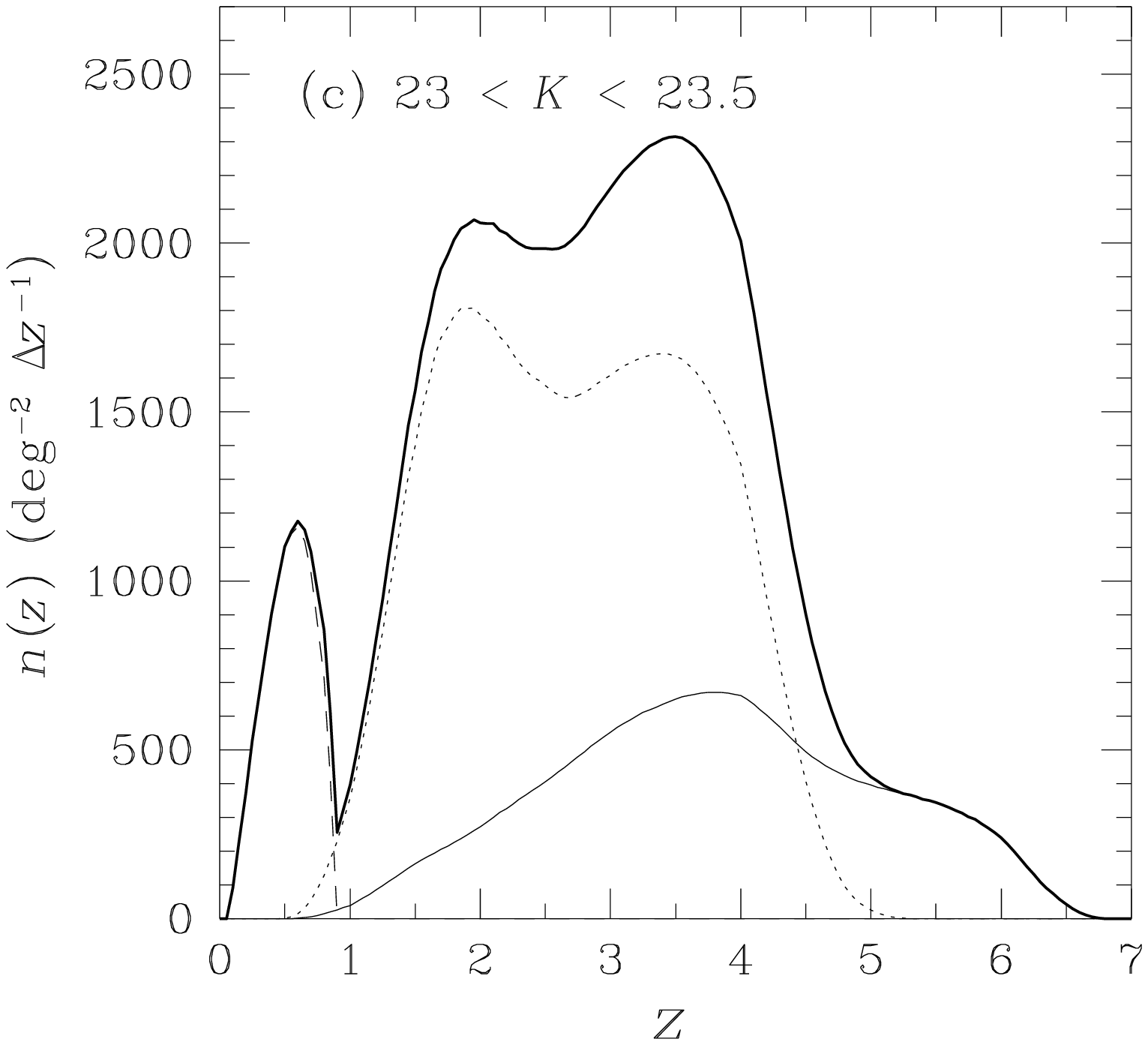}
\caption{(c) Same as (a), but for $K=23-23.5$ mag.}
\end{figure}
 
The calculation indicates that spiral galaxies dominate at lower
redshift, whereas galaxies at higher redshift are
predominantly of early-type.
For $K=17-18$ mag,
the distribution consists of spiral galaxies
with $\zmed\simeq0.4$ and early type galaxies with
$\zmed\simeq0.8$.
Although the morphologies of the $K$ selected sample are not
available, one can use emission features to show consistency 
with this prediction.
We show in Fig. 8 equivalent widths of
[OII]$\lambda3727$ emission lines for galaxies in the same magnitude
range. The figure shows a trend of decreasing equivalent widths
with increasing $z$. Though the evidence is not compelling,
these data are consistent with increasing fraction of early type
galaxies with increasing $z$, as the model calculation
predicts. This may appear to be in contrast to the 
claim that emission lines become stronger as $z$ increases.
Of course, these two statements are not contradictory: in our
case we confine ourselves to a fixed magnitude range.
A general trend of strengthening emission features to higher
$z$ is understood as  star formation being more active in spiral
disks, as in our model.
 
\begin{figure}
\plotone{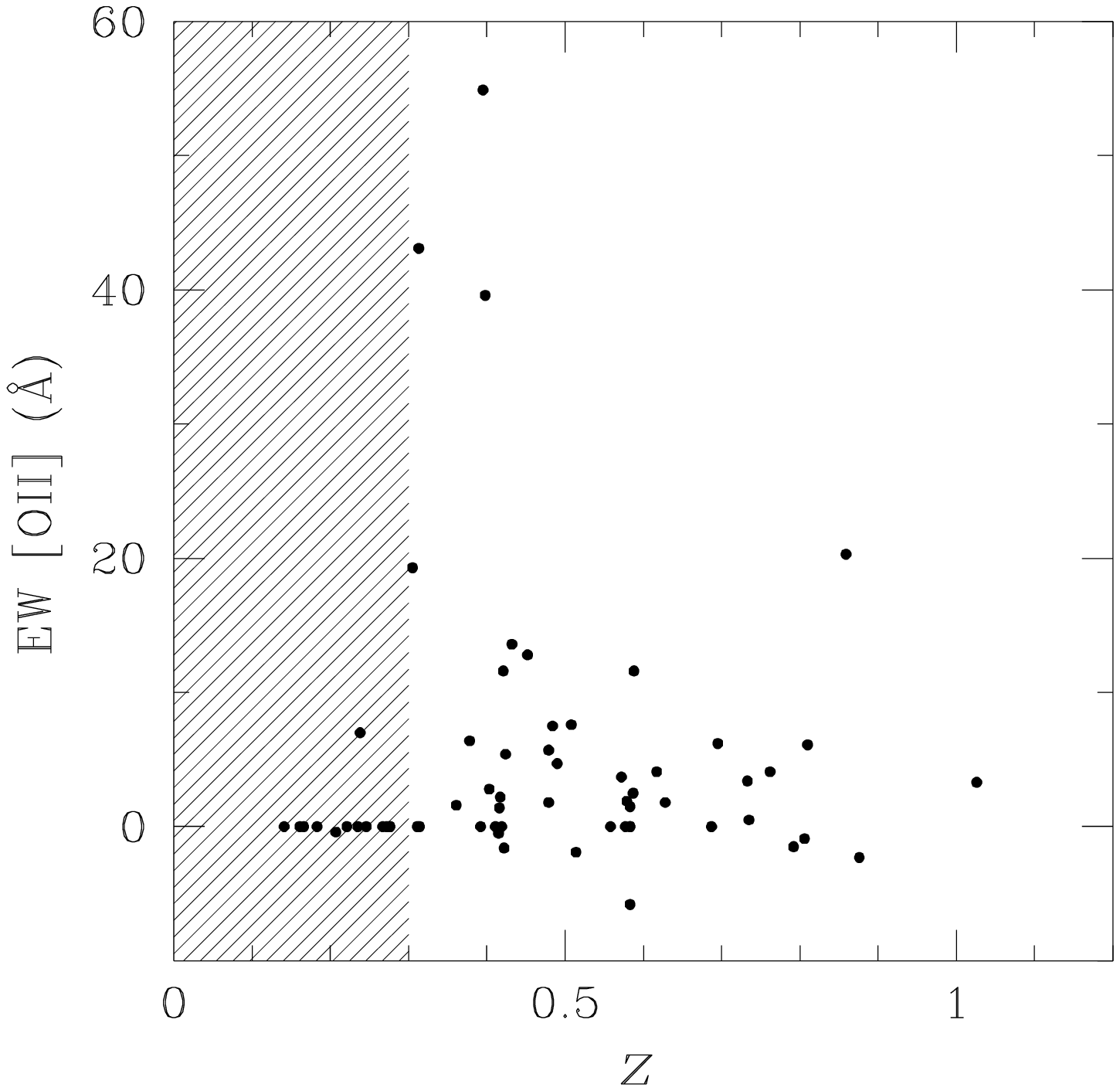}
\caption
{[O II] equivalent widths for galaxies in the
$K=17-18$ mag sample of Songaila et al. (1994), plotted as a
function of redshift.  The shade shows
the spectroscopic limit.
\label{fig8}}
\end{figure}

The maximum redshift reaches $z=1.5$ even at
$K=17-18$. This increases to $z=3.5$ at $K=19-20$ and
to $z>5$ at $K=23-23.5$.
These maximum redshifts are so high that they allow exploration of the
formation and early evolution history of spheroids, if these
high redshift tails can be sampled.
At the deepest magnitude, $K=23-23.5$, $\zmed$ of elliptical
galaxies is 3.5, although
60\% of galaxies are spiral galaxies at around $z\sim2$.
The high redshift tail of $z\gsim6$ picks up UV emission from elliptical
galaxies, and is particularly sensitive to the formation  history.
Conversely, the UV from spheroids does not contribute until
this very high $z$ is reached at the faintest magnitudes;
the prediction depends solely on optical emission,
the evolution of which is reasonably well understood within a
passive evolution model.
 
Another noticeable feature is a small fraction of
irregular galaxies, which increases only a little with
magnitude. The dominance of giant galaxies means that
the effect of evolution is very small as a whole, as often said is
consistent with no evolution, at least for $K<19$ mag
(e.g., Songaila et al. 1994).
 
If the burst-like formation epoch of elliptical
galaxies is lower than $z=5-6$,
we expect an appreciable fraction (15\%) of galaxies having
this redshift even in a $K=17-18$ mag sample (see Fig. 9). This fraction
increases to 30\% at $K=19-20$ mag. Though this looks somewhat
unusual, we are not able to exclude
this possibility immediately because of the sample
incompleteness of the redshift survey of Songaila et al.
(1994), which amounts to 15\% at $K=17-18$ mag.
 
\begin{figure}[t]
\plotone{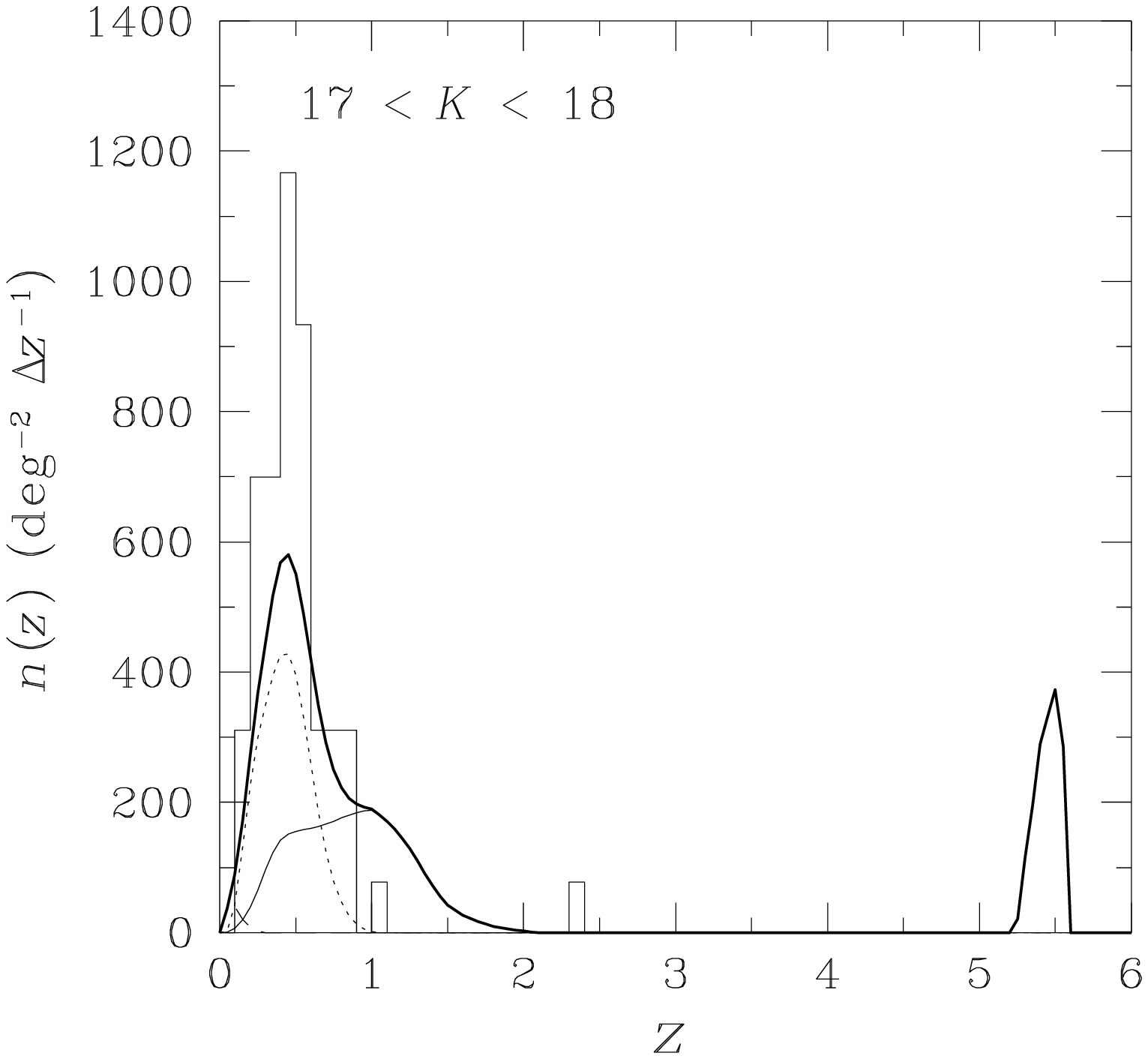}
\caption
{Same as Fig. 7 (a), but all spheroids are 
supposed to have formed
at $z\approx 5.5$ by a burst-like event.
\label{fig9}}
\end{figure}

%
%
 
\subsection{Galaxies in the $B$ band}
 
\noindent
{\it Number counts}
 
The astrophysics of the $B$ band count is very different 
from that in the $K$ band, as expected from Figure 6 shown above. 
The light, especially that from faint objects, is dominated
by young components, disks and irregular galaxies.

\begin{figure}
\plotone{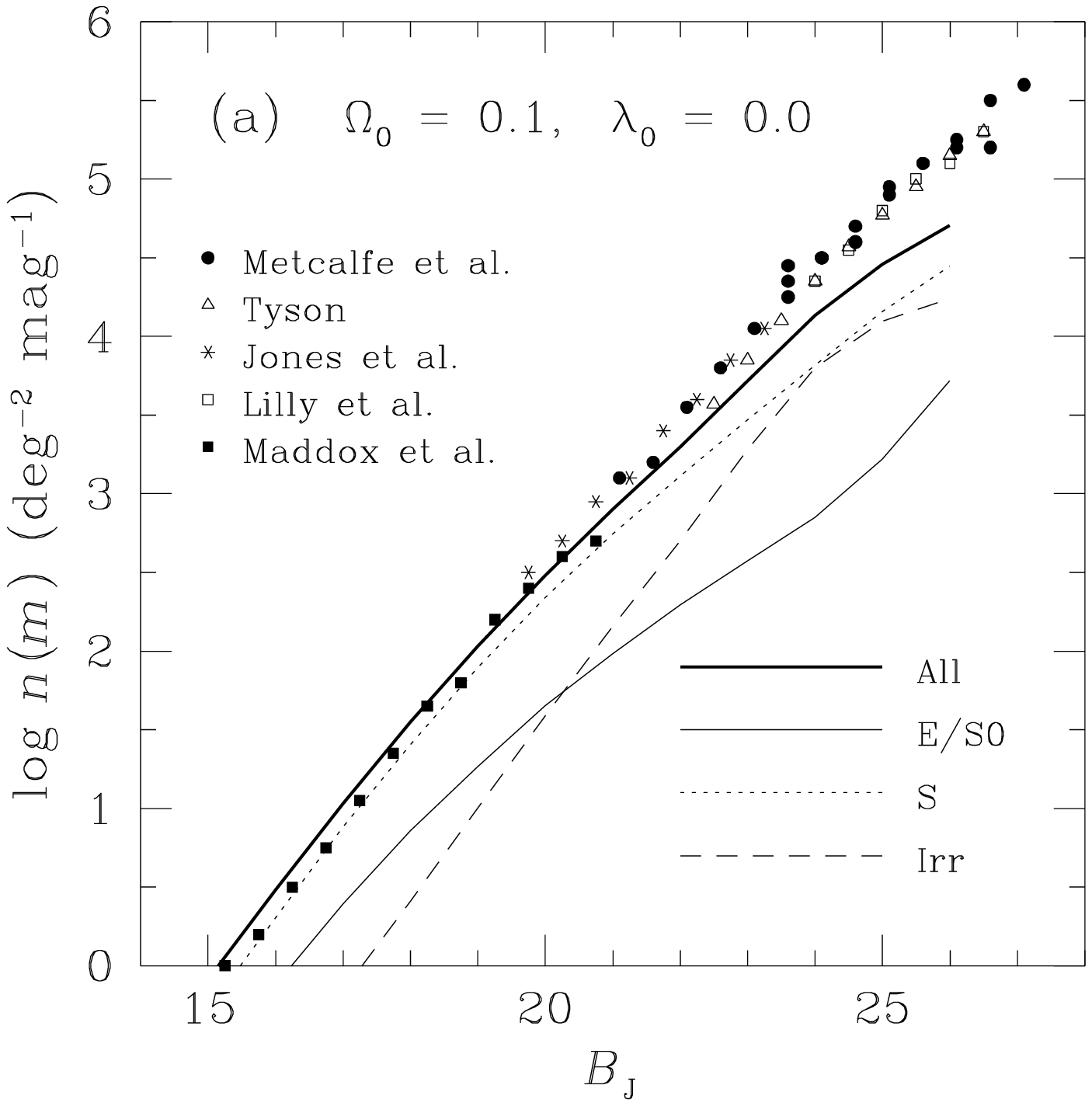}
\caption
{(a) Galaxy number count in the $\BJ$ band, as compared
with the prediction by the open cosmology ($\Omega=0.1$) model. 
Data are taken from Metcalfe et al. (1995), Tyson (1988), 
Maddox et al. (1990), Jones et al. (1991) and Lilly et al. (1991).
\label{fig10}}
\end{figure}

\setcounter{figure}{9}
\begin{figure}
\plotone{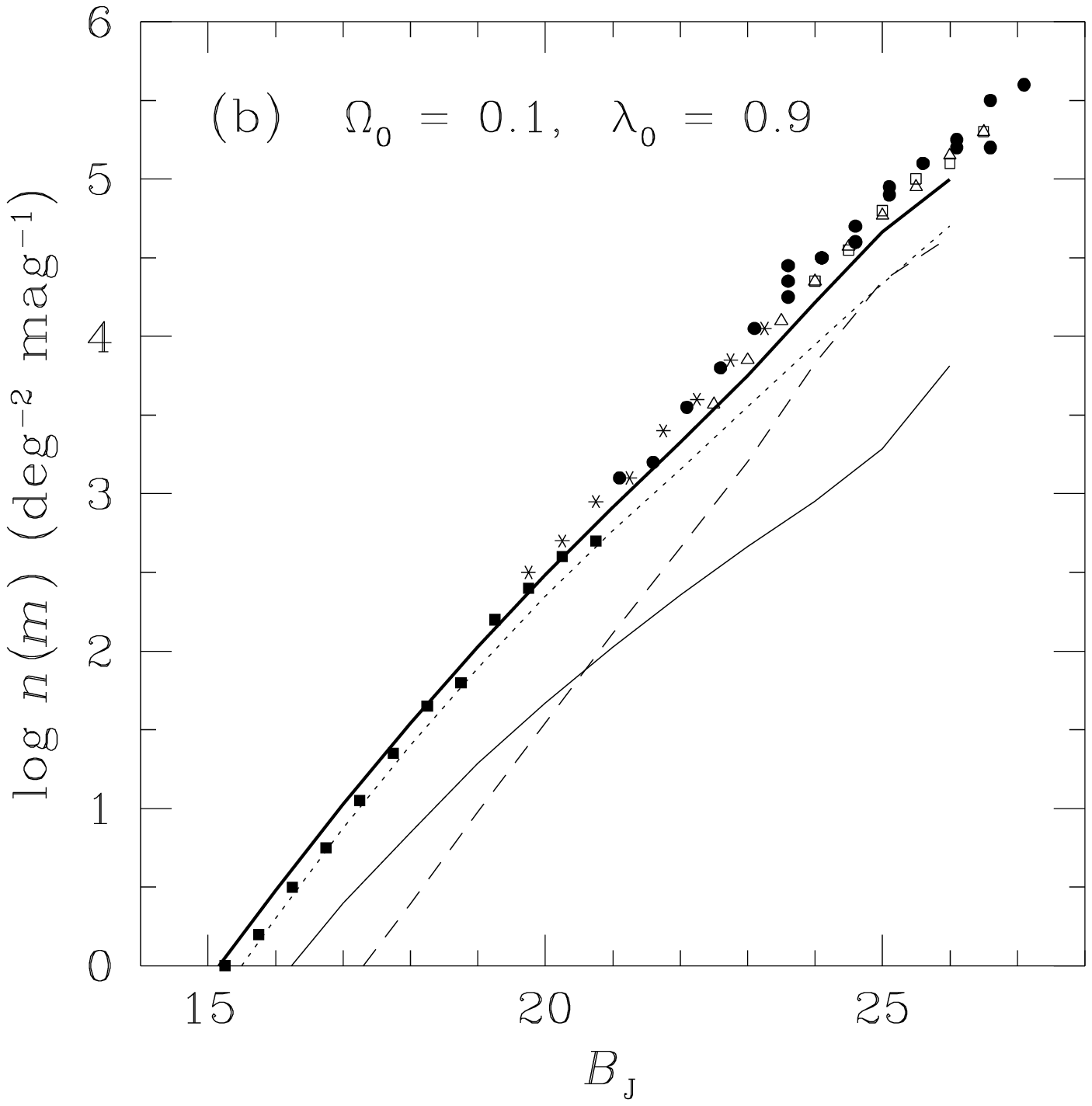}
\caption{(b) Same as (a), but for 
$\lambda$-dominated cosmology ($\Omega=0.1, \lambda=0.9$).} 
\end{figure}
 
We present in Fig. 10 the prediction of the $\BJ$ band count for
the two cosmological models:
(a) $\Omega=0.1$, and (b) ($\Omega=0.1$, $\lambda=0.9$).
The data are taken from Metcalfe et al. (1995), Tyson (1988), 
Maddox et al. (1990), Jones et al. (1991) and Lilly et al. (1991). 
We do not calculate the counts beyond $B=26$ mag (predicted
median redshift is 0.8) because of the lack
of reliable $K$ and $E$ corrections for redshift above $z=2$. 
The thick line represents the total count,
and the thin solid line and the dotted lines are E/S0 and spiral
galaxies, respectively.
The agreement between the calculation and the data is reasonable,
although the former is short of the latter by 0.3 dex at the
faintest magnitude in (a).
The contribution from spiral galaxies
remains significant even at faint magnitude, but the significance
of E/S0 galaxies diminishes for fainter magnitudes,
due to a large $K$ correction. The most
conspicuous in the figure 
is a rapid increase of Irr (dashed line) as we go to
fainter magnitudes: the contribution to the counts relative to
spiral is 10\% at 17 mag, but they contribute equally at 24 mag.
This is explained by a combination of three
effects: (i) Irr's are located at relatively near distances, and so the
reduction of spatial volume in a relativistic cosmology is only
modest; (ii) the blue-weighted
nature of spectrum gives only small $K$-corrections; and most
importantly, (iii) these galaxies formed at low redshift
and a large evolutionary effect is present.
These effects together make the slope of Irr counts quite steep.
We can conclude that these irregular galaxies are the agent that
yields a steep slope of the $B$ count.
We note that the component that gives an important contribution
at 22-25 mag is not ``dwarf'' galaxies, but is those which are
on the immediate extension of late-type spiral galaxies
to the fainter side. 
These galaxies have sub $L^*$ luminosity today,
but were nearly as bright as giant spiral galaxies
for a few Gyr after their
births.  This explains the ``faint blue galaxy problem''.
 
In the above figure, we see a better fit with a high $\lambda$
model, but we do not claim that this model is favoured.
The discrepancy between the prediction and data
seen in $\Omega=0.1$ model almost disappears, if the normalization
of the LF for Irr galaxies is increased by a factor of $\simeq$2-3,
which is within the observational uncertainties, but also
is indicated by a few analyses of local LF including blue
galaxies (Metcalfe et al. 1991; Marzke et al. 1994; Lilly et al. 1995;
SubbaRao et al. 1996)
\footnote{
The rise of LF towards faint magnitude has also been noted in the
Virgo LF (Binggeli, Sandage \& Tammann 1988).  
In this case the rise is
ascribed to that in dwarf spheroidals, but not irregulars.
It is, however, likely that irregulars evolve to dwarf
spheroids in the environment rich of galaxies, since
the gas driven out by supernovae may easily be accreted onto
giant galaxies rather than onto the original galaxies, and
irregulars may readily evolve into dwarf spheroidals.
Therefore, it is likely that dwarf spheroidals are descendants of
high $z$ irregulars and these two observations
are mutually consistent.}.
Alternatively, the possibility also exists that the LF of irregulars
is actually lifted as $z$ increases (Ellis et al. 1996), and the
normalization of the local LF for irregulars used in this paper 
is correct. 
This requires that the number of irregular galaxies decreases 
with decreasing redshift, for instance by these galaxies being 
accreted onto nearby giant galaxies.
These ``minor'' mergers modify the property of giant galaxies very
little, and this possibility seems perfectly viable, as a minimum
modification of the scenario presented in this paper.  On the other
hand, it is unlikely that these population has simply faded away.
 
BES have discovered that [OII] emission increases rapidly with 
increasing magnitude. The fraction of galaxies that show
equivalent widths larger than 20\AA~ increases from 15\% at $\BJ
\le17$ mag to as much as 55\% at $\BJ\sim21$ mag. Koo \& Kron (1992)
ascribed this increase to the increase of the fraction of
subluminous late type galaxies, which generally are strong [OII] 
emitters, as one goes to fainter magnitude. 
We agree that this is one cause, but not the major one.
In our model, this increase is caused by both an increase 
of Irr fractions
as a result of a steep slope of $N(m)$ for Irr, {\it and} an
increase of gas in spiral galaxies.  Straightforward reading of
the [OII] equivalent width distribution of
Kennicutt (1992) (see his Figure 9)
gives 13\% of normal galaxies having $>20$\AA~
[OII] equivalent widths, consistent with the BES $\BJ<17$ mag sample.
The disks at $z\sim 0.3-0.4$ are two times more gaseous than at
$z=0$. A simple calculation using the [O II] distribution
in the Kennicutt figure leads to the result that the fraction 
of galaxies having EW([OII])$>$20\AA~ is 45\% at $\BJ\sim21$. 
This is consistent with the result of BES.
 
In our explanation of the B band count, we do not particularly invoke
incidental star burst activity.  Star formation takes place steadily
as it does today, but those galaxies at $z\sim0.5-1$ are much more
gaseous: the gas in Irr 1 Gyr after their birth is 3 times that
of today, and so the global star formation rate per galaxy is 3
times higher, which might give the appearance that those galaxies
are undergoing star bursting activity.
 
\begin{figure}[t]
\plotone{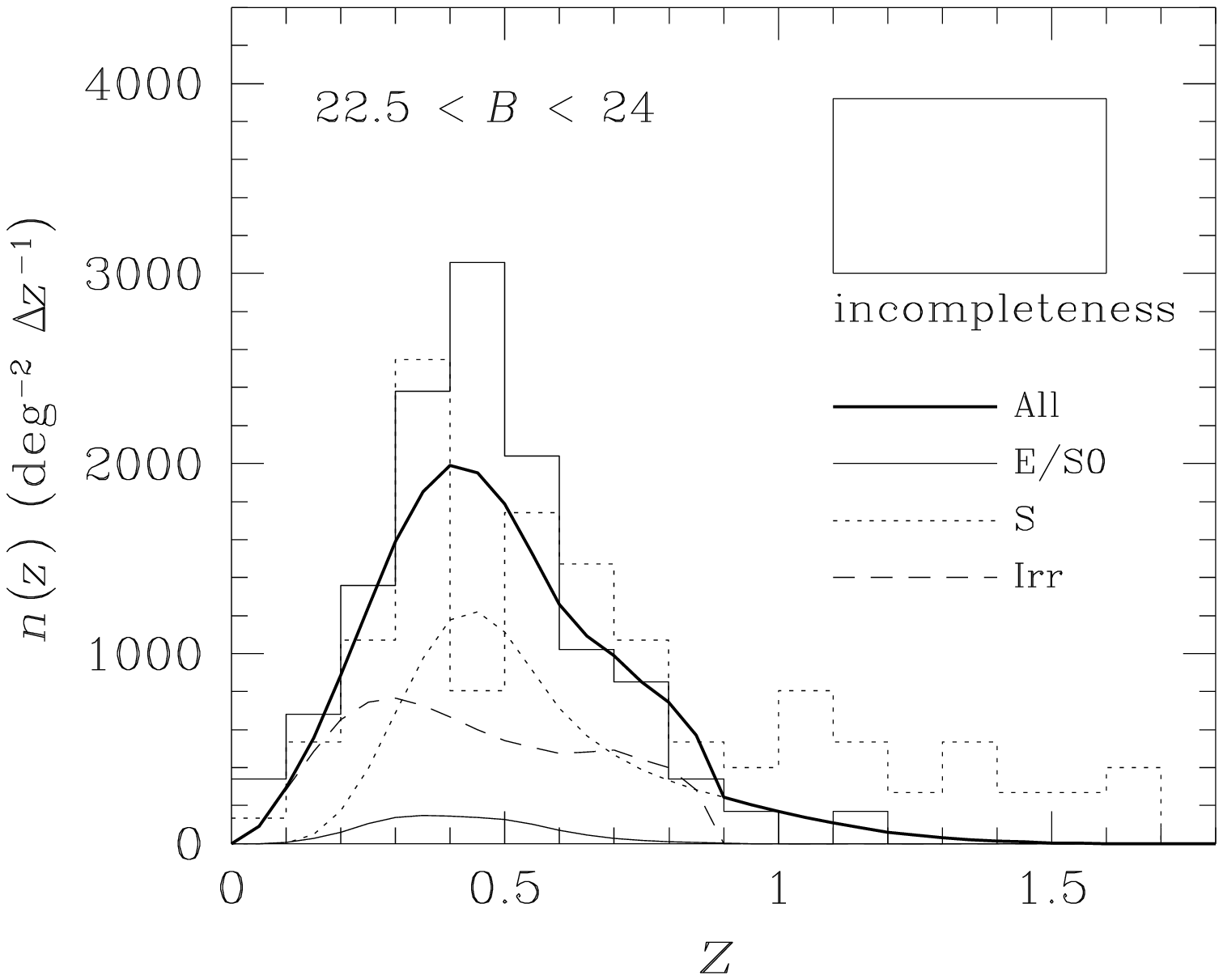}
\caption
{Redshift distribution of galaxies selected with
$\BJ=22.5-24$ mag, as compared with the model prediction. 
Data are taken from Glazebrook et al. (1995a; solid histogram) 
and Cowie et al. (1996; dotted histogram), 
and normalized to the count data.
\label{fig11}}
\end{figure}
 
\smallskip

\noindent
{\it Redshift distribution}
 
The redshift distribution of the sample selected in the $B$ band
corroborates the view given above. In Fig. 11, we compare the
prediction with the observations of Glazebrook et al. 
(1995a; solid histogram) and of Cowie et al. (1996; dotted histogram) 
for $\BJ=22.5- 24$ mag. 
We can see a good agreement between the two for the dominant
feature around the peak. 
The redshift data are normalized to the observed count.
An interesting fact is that the observed distribution looks quite 
similar to that of spiral (plus E/S0) galaxies, 
while the normalization of
the former is higher by a factor of two. This is the feature
that has puzzled many cosmologists after the discovery by
British groups (BES; Colless et al. 1990).
In our model this factor two difference
is explained by irregular galaxies which were almost
as luminous as giant spiral galaxies in the blue band.
The survey selected in the blue light preferentially samples
actively star forming components, which were newly formed in the
near past.  At this magnitude the contribution from E/S0 is
quite small.
 
The presence of high $z$ tail has been a matter of debate. Earlier
observations  (BES; Colless et al. 1990; Lilly, Cowie \& Gardner 1991;
Glazebrook et al. 1995a) have claimed the absence of the high $z$ tail.
On the contrary, some recent observations (Cowie et al. 1996; Koo 1996)
indicate the presence of the tail. Our model predicts a small high 
redshift tail which arises from luminous spiral galaxies; 
we cannot eliminate entirely the high $z$ tail. The size of tail 
depends largely on the amount of evolution of spiral galaxies. A large
high $z$ tail of Cowie et al., if confirmed, suggests that the evolution
should be stronger than is in the present model.
 
%
%

\begin{figure}[t]
\plotone{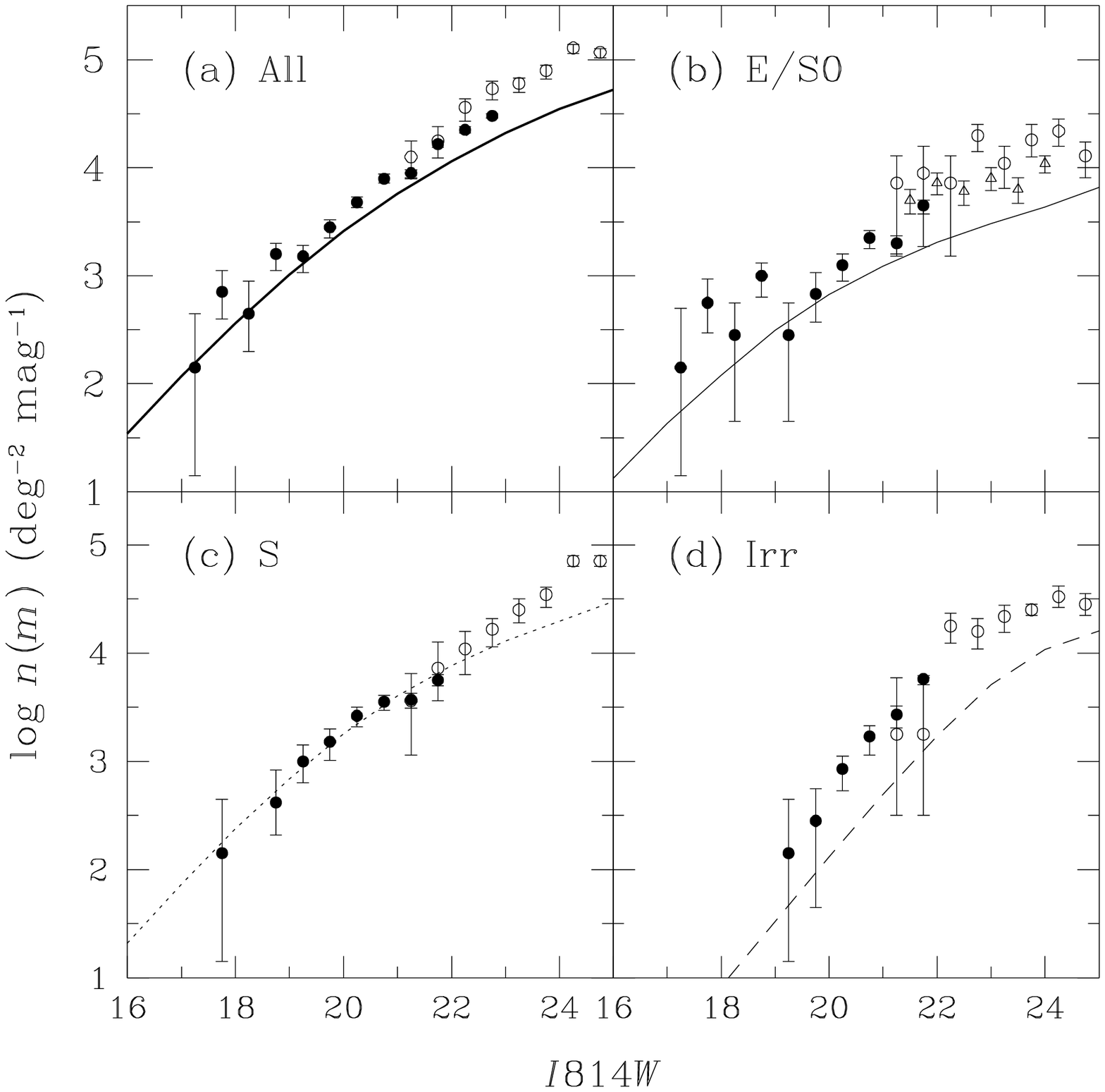}
\caption
{Galaxy number count of morphologically classified
sample, compared with the model prediction for an open universe
$\Omega=0.1$.  Data are taken from Glazebrook et al. (1995b)
(solid circles), Abraham et al. (1996) (open circles) and
Driver et al. (1996) (open triangles, for (b) only).
\label{fig12}}
\end{figure}

\begin{figure}[t]
\plotone{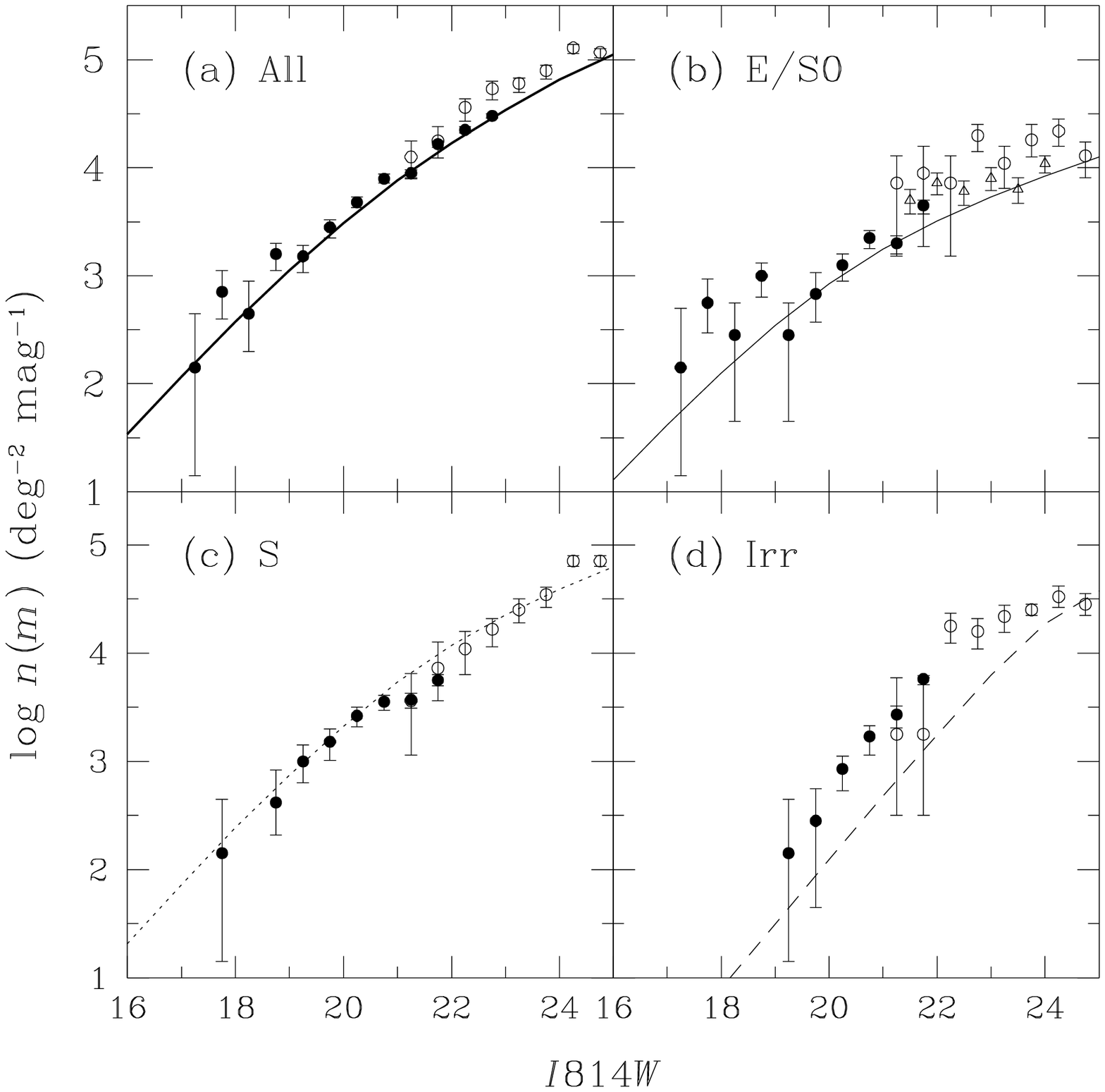}
\caption
{Same as Fig. 12, but for the $(\Omega=0.1, \lambda=0.9)$ model.
\label{fig13}}
\end{figure}

\subsection{Galaxies in the $I$ band}
 
\noindent
{\it Number count}
 
The novel feature for the $I$ band is that morphologically classified
number counts are available (Glazebrook et al. 1995b; Driver et al.
1996;
Abraham et al. 1996) to $I=24$ mag.
At this magnitude we expect the median redshift to be unity.
We compare the prediction of the $\Omega=0.1$ model 
with the observation in Fig. 12
for E/S0 (panel b), spiral (panel c), Irr (panel d) and the 
total sample (panel a).   
We see a good agreement for spiral galaxies to $\I814=22$ mag.
For E/S0 galaxies the predicted curve is
somewhat lower by $\approx 0.1-0.2$ dex to the same magnitude,
but the statistics are poor for this type.
The disagreement is significant for Irr.
The important feature here, however, is that the
steep slope of the Irr counts (index=$0.64\pm0.05$ for
$I=19-22$ mag) is correctly reproduced by a sub $L^*$ population of
young, quickly evolving galaxies.
That the predicted counts of irregulars are lower by a factor of three 
is similar to the problem we encountered with the
blue count, and can be accounted for if the LF of
Irr is increased by a factor of 3. 

Glazebrook et al. (1995b) have noted that the normalization of their
$I$ count is higher than predicted from the local LF in the $B$ band. 
In our calculation this discrepancy is not so conspicuous: 
a half of the discrepancy is accounted for
by our higher normalization of the local LF, which is larger than
that of Loveday et al. (1992) by 0.15 dex. We still see a
discrepancy for the amount of 0.12 dex at $\I814=22$. This
is reduced to 0.05 dex if the normalization of the Irr-LF is
multiplied by a factor of 3.

At fainter magnitudes  the observed counts
are somewhat higher than calculated. This is particularly true for 
E/S0 galaxies. We note that the shape of the prediction of E/S0
galaxies is rather generic and it is not easy to modify the prediction.
On the other hand, somewhat higher observed counts of spiral galaxies
may be explained by increasing evolution of spiral galaxies. It may
also be possible that there is a misclassification between spirals
and Irr galaxies, since the classification becomes increasingly 
ambiguous towards the faintest magnitude: the number count of Irr 
may not flatten but increase linearly towards the fainter magnitude, 
while the number of spiral galaxies may rise close to the predicted 
curve. It is also possible that the counts from  $I>22$ mag survey are
largely affected by large-scale inhomogeneity of the galaxy 
distribution, since the survey fields are quite small.

Driver et al. (1996) concluded that
high $\lambda$ models ($\lambda\ge 0.7$ in a flat cosmology)
are excluded from a comparison of their
E/S0 counts with their model prediction,
as the predicted count with high $\lambda$ overshoots the observation.
Our model calculation, however, does not confirm their
conclusion. We present in Fig. 13 the predicted counts for
$(\Omega=0.1, \lambda=0.9$) model, where the discrepancy 
between the prediction and the data almost disappears to the
faintest magnitude. In particular, the prediction of E/S0 is
consistent with data of Driver et al. We note that
the main difference between our model and Driver et al.'s
lies in the normalization of E/S0 galaxy counts, i.e.,
a much higher normalization adopted by the latter work. 

\begin{figure}[t]
\plotone{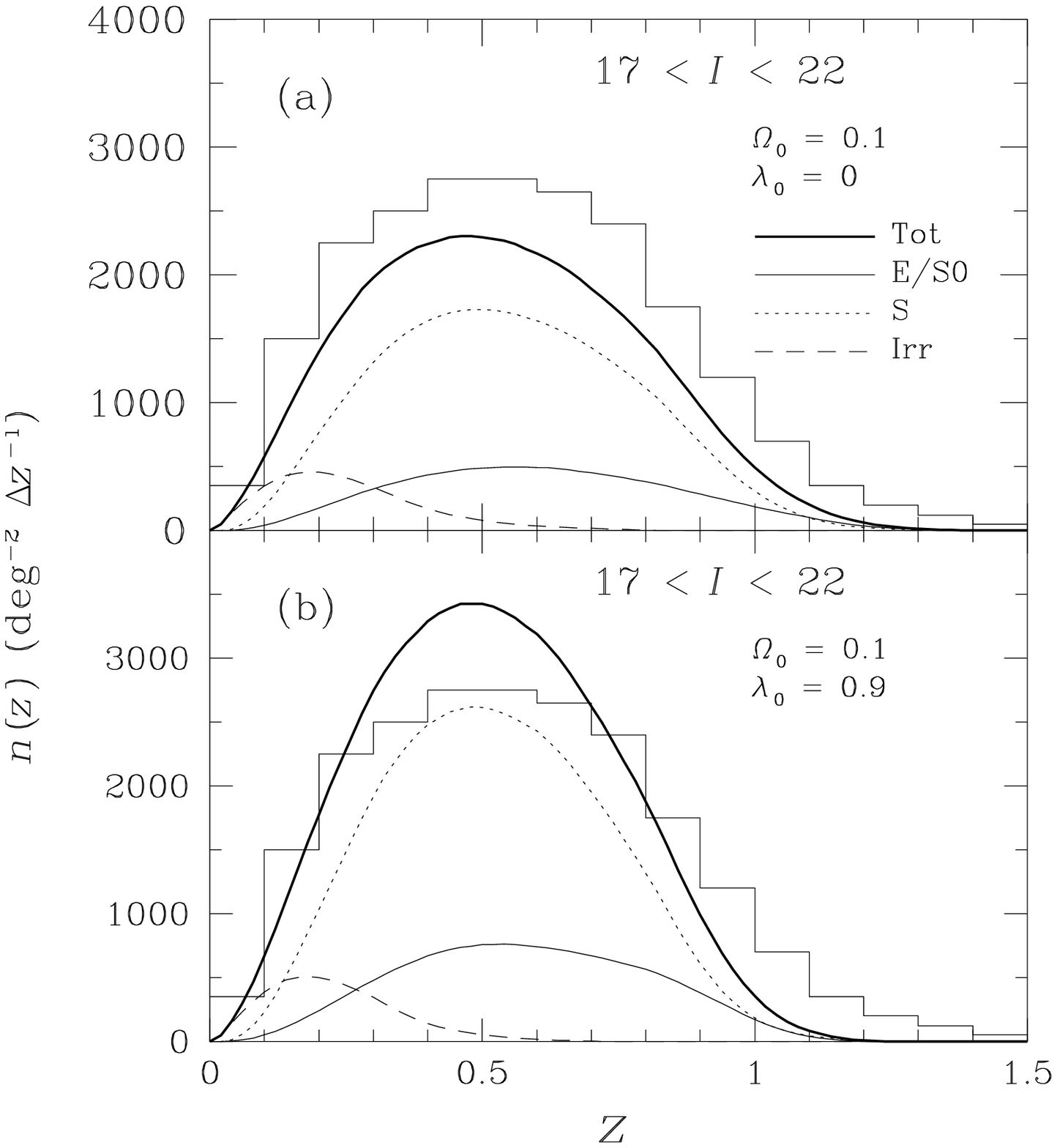}
\caption
{Redshift distribution of $I$ band selected sample
for $I=17-22$ mag. The prediction is compared with the 
observation of Lilly et al. (1995).  (a)  open
cosmology $\Omega=0.1$; (b) $\lambda$-dominated cosmology ($\Omega=0.1,
\lambda=0.9$).
\label{fig14}}
\end{figure}

\begin{figure}[t]
\plotone{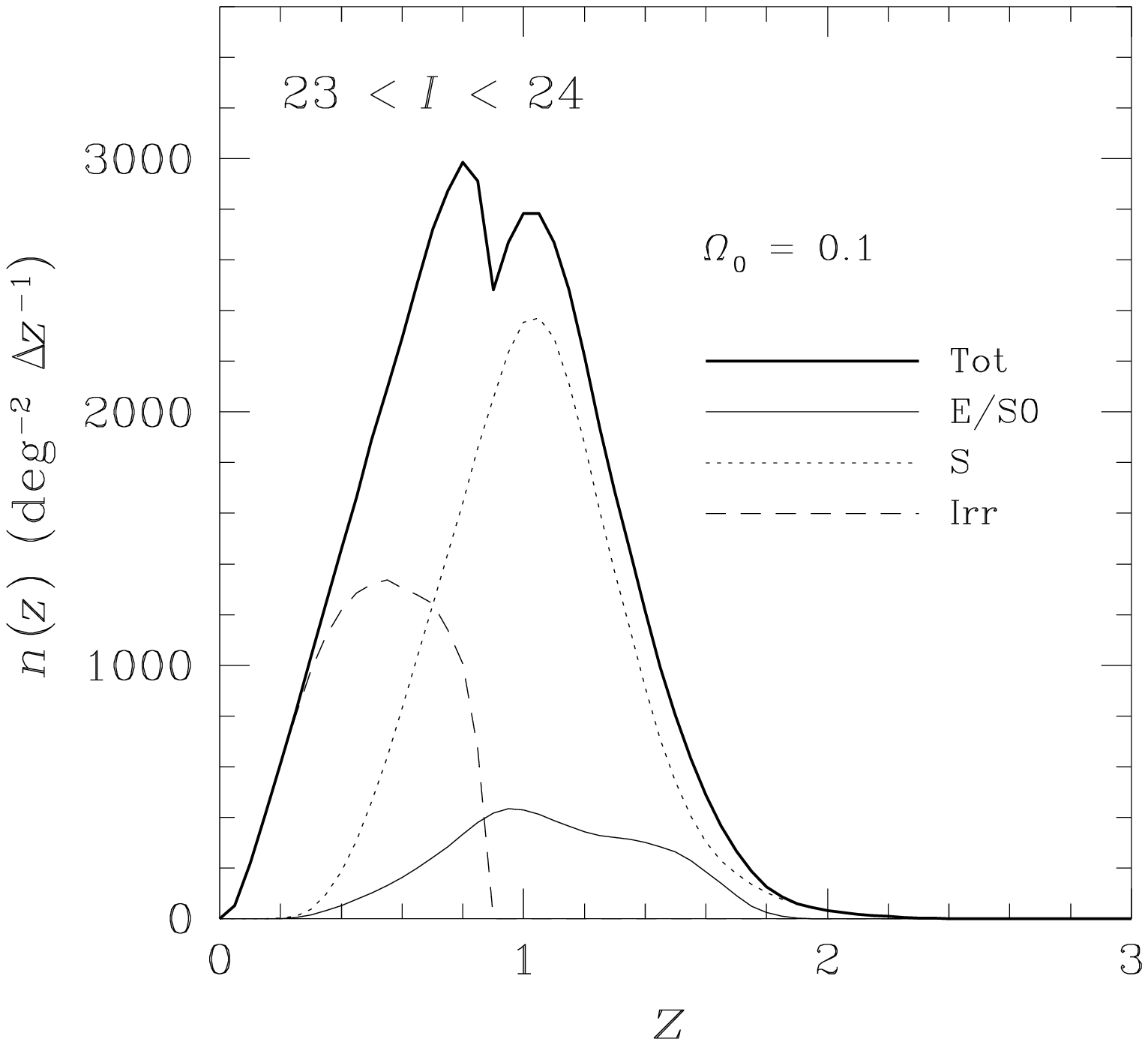}
\caption
{Prediction of the redshift distribution for the
$I=23-24$ mag sample for the open universe model.
\label{fig15}}
\end{figure}

\smallskip
\noindent
{\it Redshift distribution}
 
Lilly et al. (1995) have carried out a large redshift survey
to $I$=22. Their data are compared with the prediction in
Fig. 14 (a). The predicted redshift distribution
agrees with the data very well, with
the normalization, however, smaller by 20\%
for the open cosmology. The case of $\lambda$ dominated
universe $(\Omega=0.1, \lambda=0.9)$ is displayed in Fig. 14 (b). 
Unlike the case for the $B$ band, the distribution is dominated by
spiral galaxies, and the contribution from E/S0 and Irr is quite
small. The small contribution from Irr is understood by the fact
that the very active star formation for $0.5<z<1$ largely affect
the $B$ band observations, whereas $I$ band observations are  more
sensitive to the total number of stars in the galaxy, and  
stars are not yet accumulated before $z=0.5$. 
We note that the evolution of spiral galaxies is less than
0.5 mag to $z=1$, and the evolutionary effect is not manifest in
the $z$ distribution. 

The prediction for the redshift distribution is shown in
Fig. 15 for $I=23-24$ mag. We note that E/S0
counts are always lower than the contribution from
spiral galaxies for all $z$.  This is different from what we saw
for the $K$ counts, where the high redshift
part was dominated by early type galaxies.
 
%
%

\begin{figure}[t]
\plotone{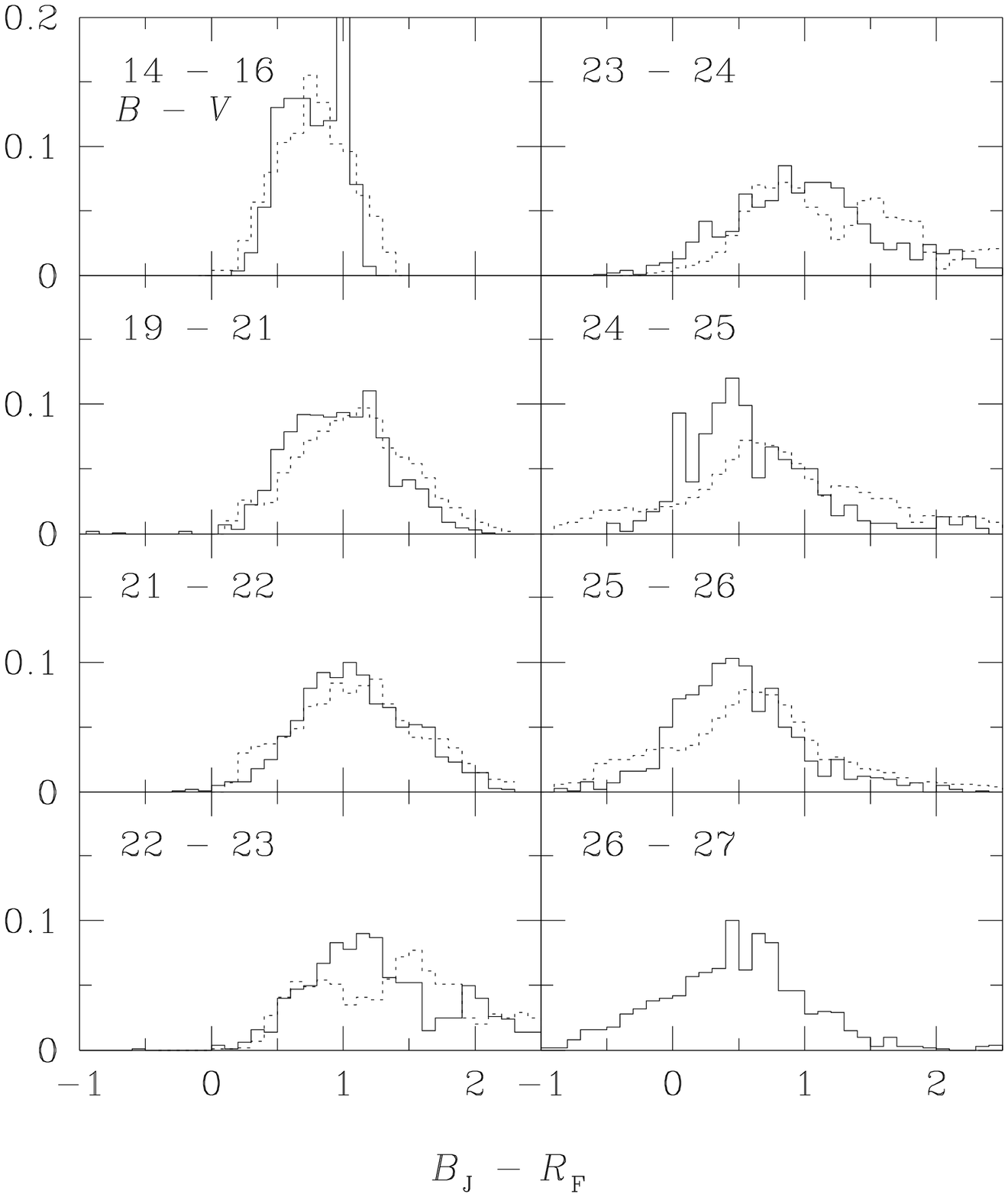}
\caption
{Colour distribution of galaxies which fall 
in the bins specified by
$\BJ$ magnitude.  The figure for the brightest bin is $B-V$,
and the rest is $\BJ-R_{\rm F}$.  The model predictions are
shown by dotted lines. The data are taken from Gronwal \& Koo 
(1995) for $B-V$ and from Koo \& Kron (1992) for the rest. 
\label{fig16}}
\end{figure}
 
\subsection{Colour distribution}
 
The colour distribution is shown in Fig. 16. We
take the $\BJ-R_{\rm F}$ data from a compilation
of Koo \& Kron (1992) for the
magnitude between $\BJ=19$ and 27. The data for the brightest
magnitudes ($\BJ=14-16$) are $B-V$ from Gronwall \& Koo (1995).
The prediction, denoted with dotted lines, agrees well with
the data to $B=24$ mag. We see that the data are slightly
bluer for fainter magnitudes. Better agreement is
achieved by increasing the irregular population
by a factor of 2, although the prediction for $\BJ=25-26$ mag is still
somewhat (0.2 mag) redder.  We do not calculate the colour
distribution for $\BJ>26$ for the reason given in section 4.2 above.
 
\smallskip
\noindent
{\bf 4.5 The Mg II absorption-line selected sample and the constraint
  on the luminosity evolution}
\smallskip
 
Steidel et al. (1994) have studied properties of
58 galaxies that yield Mg II absorption lines near the line of
sight to quasars. Giant galaxies are always found close
to the quasar sightline along which Mg II absorption lines are
detected, and conversely all giant galaxies located close to
the quasar sightline cause Mg II absorption in the quasar
spectrum. Their conclusion is that those galaxies
that cause Mg II absorption lines are giant galaxies and these
galaxies hardly evolve at least between $z=0.3$ and 1.0:
the allowed amount of evolution is 0.1$\pm 0.2$ mag in the
rest frame $B$ band for this redshift interval
\footnote{See Schade et al. (1996), however.}. 
This puts a very strong constraint on the model. 
In our model, E/S0 undergo
evolution by a half magnitude and Sc galaxies by 0.2 mag
(see Fig. 17). While our evolution of E/S0 galaxies looks 
faster, the fraction of galaxies that show E/S0 colours is small
in the Steidel et al. sample, and their data do not give a
strong constraint on the evolution of E/S0 galaxies.
Irregular
galaxies also evolve faster, but such galaxies do not yield Mg II
absorption lines and are excluded from the sample; hence no 
constraint is obtained on the evolution of irregulars.
Therefore, the prediction of our model is consistent with the result
from the Mg II sample.

\begin{figure}[t]
\plotone{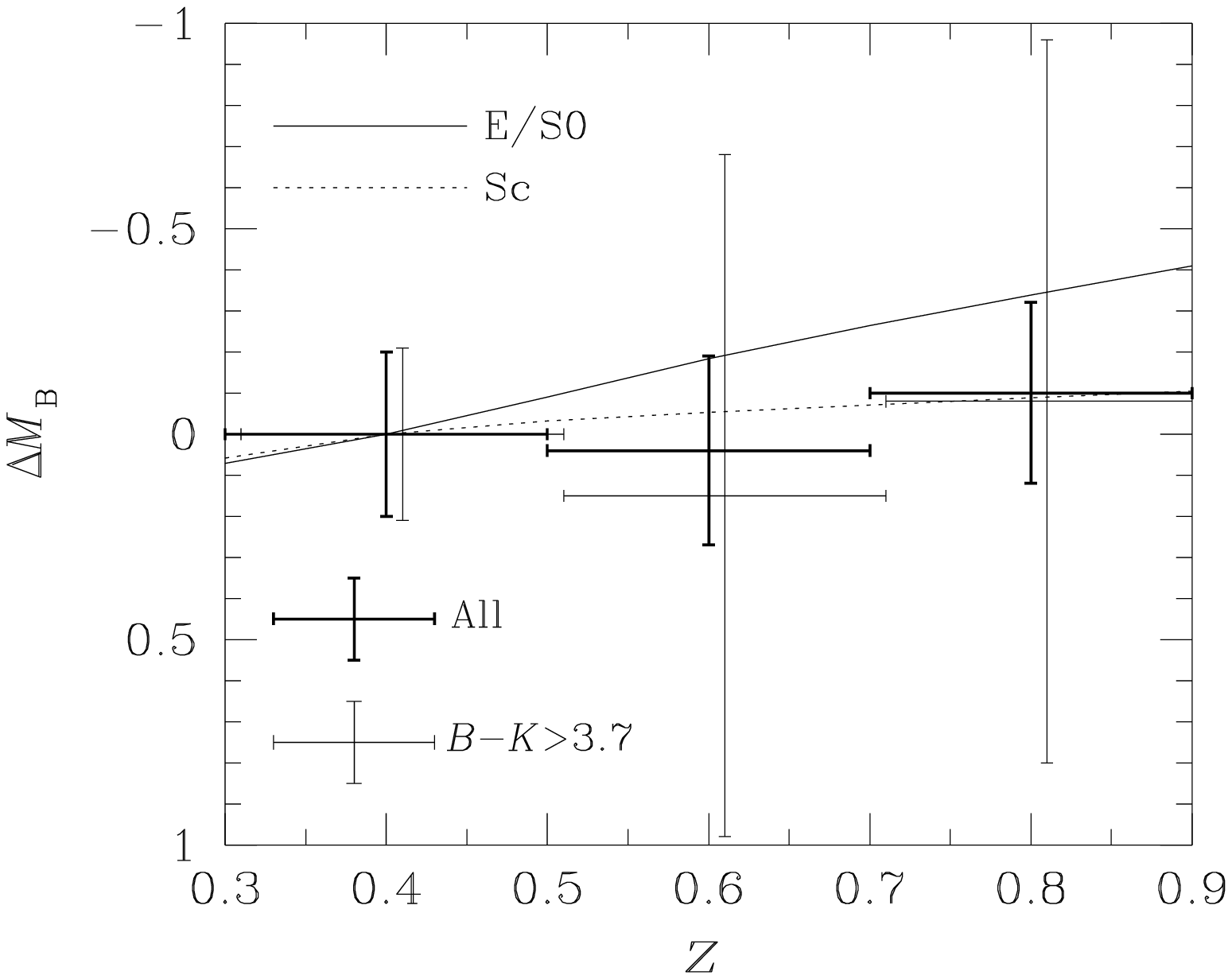}
\caption
{$B$ band luminosity of giant galaxies in the
Mg II quasar absorption line sample of Steidel et al. (1995).
Data points denoted by thick lines represent the total sample,
and those by thin lines show the sample of red galaxies with 
$B-K>3.7$, nominally corresponding to elliptical galaxies.
Two curves show the prediction for E/S0 and Sc galaxies
with their luminosity normalized at $z=0.4$.  
\label{fig17}}
\end{figure}
 
We note that Sc galaxies undergo evolution, after gas infall
has ceased, by 0.4 mag between
$z=0$ and $z=0.3$, which is out of the range surveyed by Mg II
absorption lines. It is interesting to note that this
causes brightening of $L^*$ below $z=0.3$, and hence increases 
the number of super $L^*$ galaxies observed at higher redshift 
by a factor of 1.7 times.
Sc is the median colour of Mg II absorbers, and
this brightening may account for
the offset in the normalization (by a factor of 1.5, if the 
normalization of the local LF used in the present paper is adopted)
between the Mg II sample and the local LF as found by Steidel et al.
 
%
%

\begin{figure}[t]
\plotone{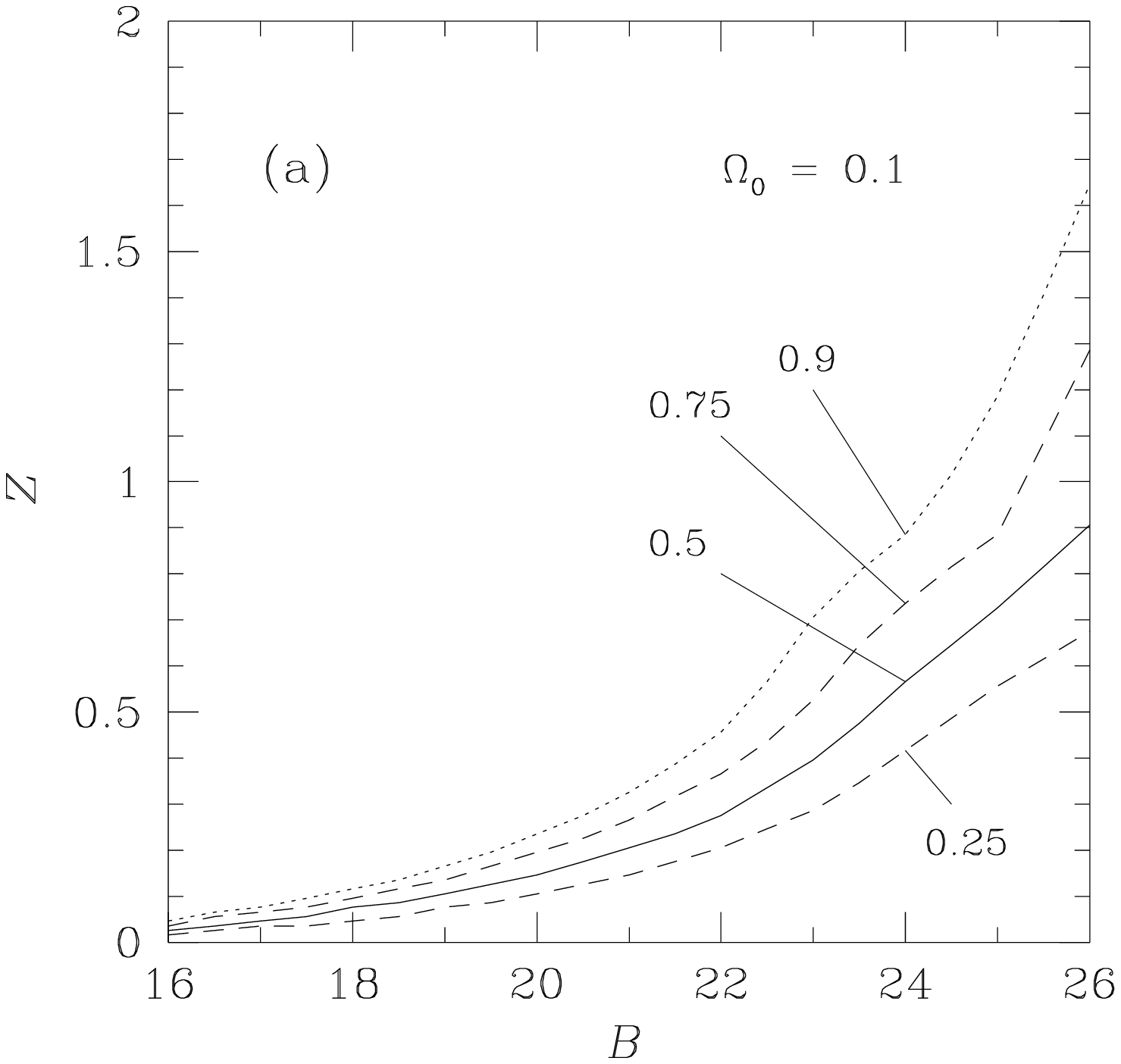}
\caption
{(a) Median (solid curves), quartile (dashed curves)
and 90 percentile (dotted curves) redshifts predicted in the model
as a function of magnitude for $B$ selected samples.
\label{fig18}}
\end{figure}

\setcounter{figure}{17}
\begin{figure}[t]
\plotone{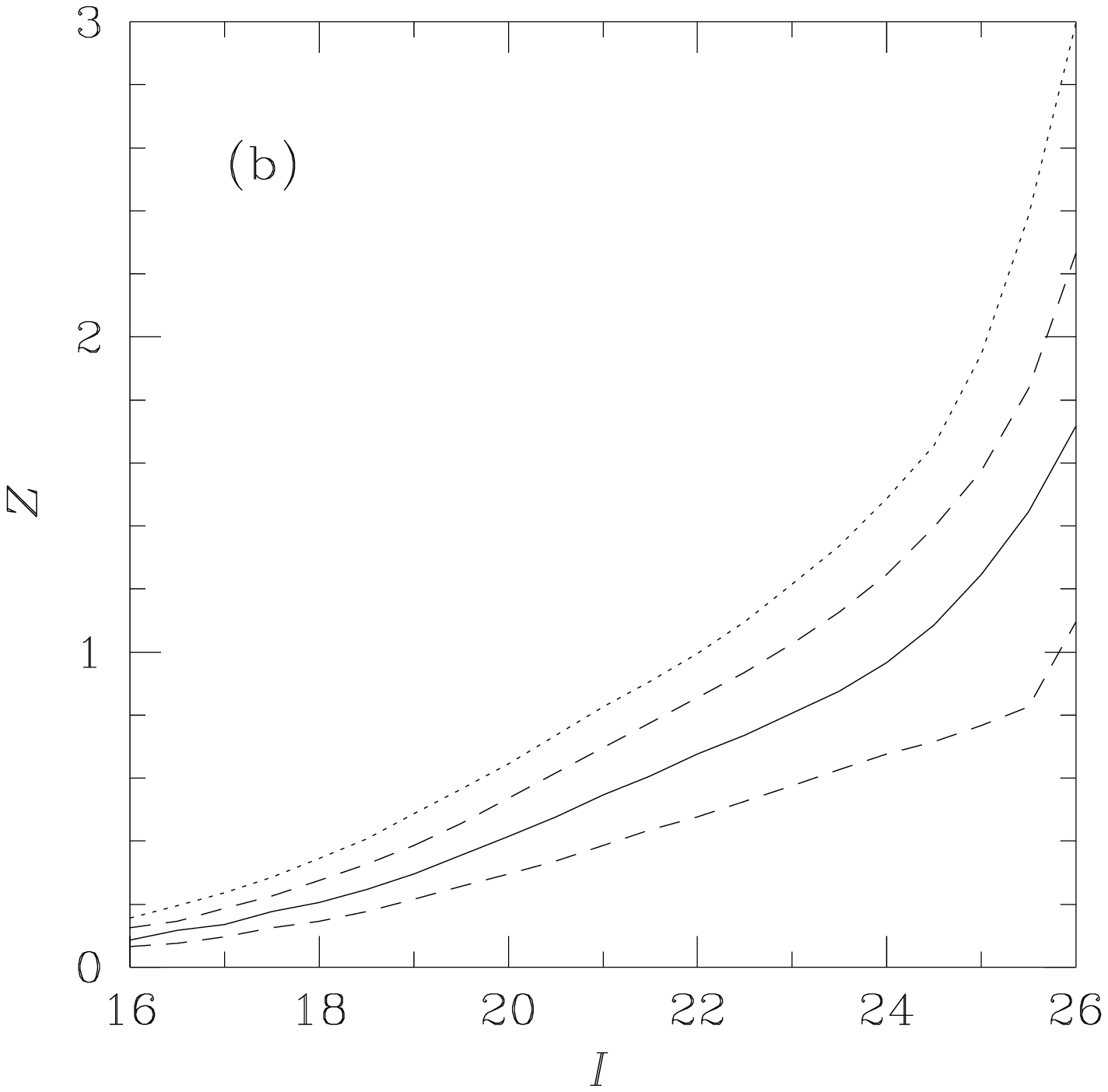}
\caption{(b) Same as (a), but for $I$ selected samples.}
\end{figure}

\setcounter{figure}{17}
\begin{figure}[t]
\plotone{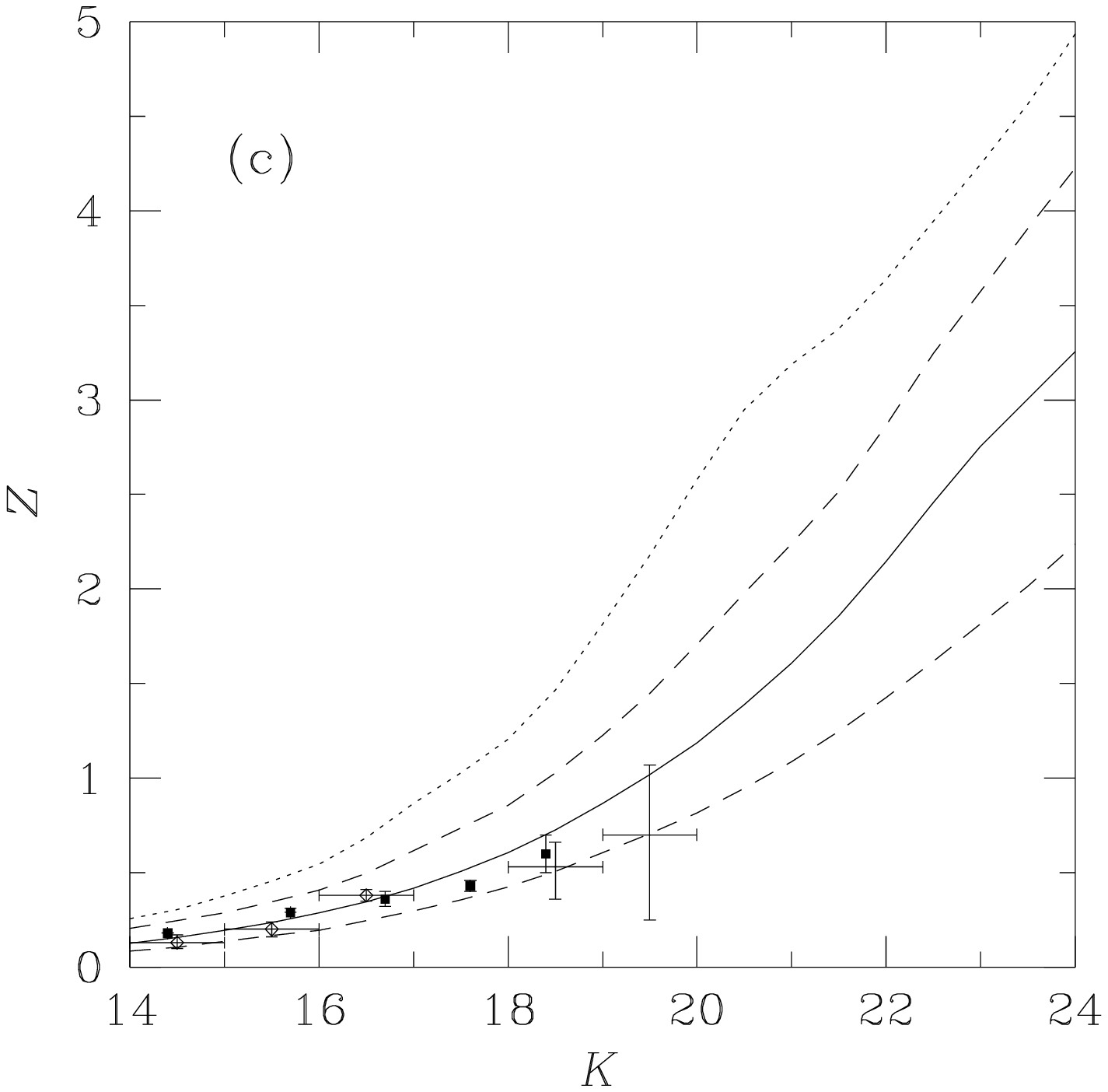}
\caption{(c) Same as (a), but for $K$ selected samples.
The median redshifts of the Cowie et al. (1996) sample are shown.}
\end{figure}
 
\subsection{Magnitude-redshift relation}
 
In order to summarize the basic prediction of our model, we present
in Fig. 18 the redshift as a function of magnitude for $B$, $I$ and
$K$ colours for the $\Omega=0.1$ model. The plotted curves are median
(solid curve), quartile (dashed curve) and 90 percentile (dotted curve).
 
%
%

\begin{figure}[t]
\plotone{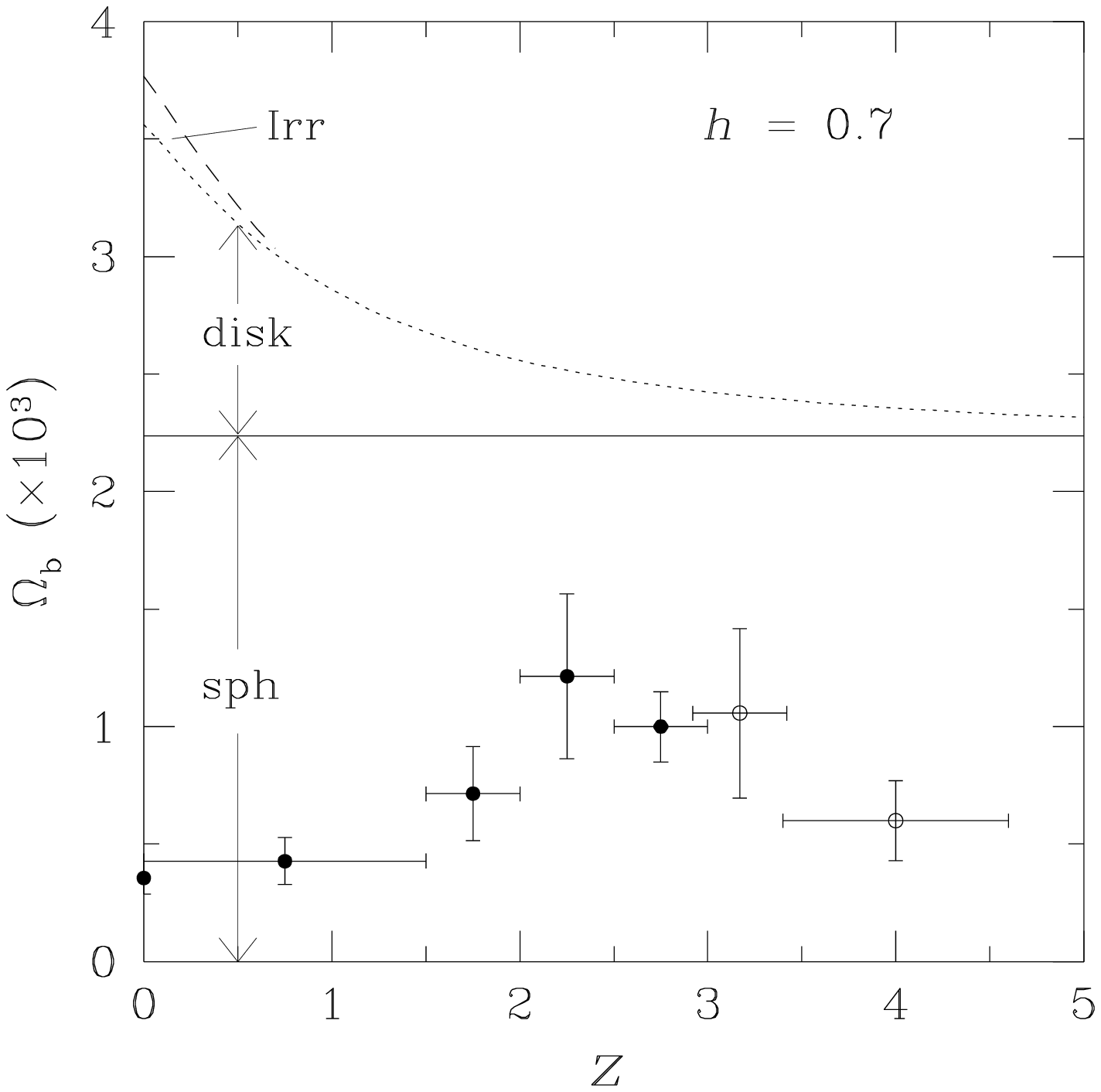}
\caption
{Evolution of baryons frozen into stras.  The three
curves represent spheroids (E/S0 + bulges of spirals), disks of spirals
and irregulars. The data show the evolution of the
neutral hydrogen gas abundance 
inferred from quasar absorption line surveys: solid circles are
from Lanzetta et al. (1995), and open circles are from Storrie-Lomnardi
et al. (1996).  The point at $z=0$ is from an HI survey of Rao \& Briggs
(1993). 
\label{fig19}}
\end{figure}

\subsection{Baryons in stars and the gas}
 
We calculate the amount of baryons frozen into stars as
a function of look-back time in Fig. 19.  The solid, dotted
and dashed curves indicate spheroids,
spheroids + disks, and total = spheroids + disks + irregulars,
respectively. The mass of spheroids is constant in this redshift
range as assumed.
The mass in disks increases to lower redshift.
The apparent rapid increase towards $z=0$
in this figure, however, is
an artefact of the compression of the time scale of the
abscissa towards $z=0$; the increase close
to $z=0$ is actually quite slow if
the diagram is represented as a function of time rather than
redshift. 

The prediction of  $\Ms/L_B$ for the spheroids with 
the population synthesis model
is far from robust, since 
it depends significantly on the lower mass cutoff, even if
the initial mass function is fixed. Charlot, Worthey \& Bressan (1996)
have given a compilation of $\Ms/L_V$ for widely accepted models,
which give $\Ms/L_V=6-8$ for the age 13-15 Gyr. 
We adopt $\Ms/L_B=8.4$, which agrees with
the value obtained from kinematics of elliptical galaxies
by van der Marel (1991), if $h=0.7$.  
 
We note that about 2/3 of baryons in stars today are in spheroids, 
and only 1/3 in disks. This agrees with the estimates obtained with
different assumptions (Fukugita et al. 1996; 
see also Persic \& Salucci 1992).
It is interesting to note that the contribution
from irregulars to the stellar mass density is very small
in spite of their drastic contribution to the $\BJ$ counts.
We note that $M(\Ms/L_B)=3.1$ and 1.8 for
disks and irregular galaxies in our model. $\Ms/L_B$
for galaxies of various morphology types are given in Table 2 above. 
 
We also plot in Fig. 19 the evolution of the neutral
atomic gas density
in units of the critical mass density obtained from surveys of
damped Lyman $\alpha$ systems 
(Wolfe et al. 1986; Lanzetta et al. 1995). We
adopt for $z\ge 3$ the data of Storrie-Lombardi et al. (1996), which
update those of Lanzetta et al. (1995): the data at high
$z$ have gone down by a factor of two, and this is ascribed to
small statistics and an underestimate of statistical errors
of the Lanzetta et al. sample
\footnote{We thank Richard McMahon for clarifying this point.}.
The point at $z=0$ is taken from an HI survey of
Rao \& Briggs (1993) (see also Fall \&
Pei 1993).  It is interesting to observe that the amount of
decrement in gas from $z=3$ to $z=0$ is just consistent with
the increment of baryons in disk stars in the same redshift range.
This is somewhat different from the claim by Lanzetta et al.
(1995) who regarded the consumption of neutral gas as being
used to form all stars in the universe.  Accepting the new result
of Storrie-Lombardi et al., the baryon budget calculated here
indicates that the decrement of the neutral gas is significantly
smaller (by a factor of 3) than the total amount of baryons 
frozen in stars. This favours the picture that most spheroids existed
prior to $z=3$, and the neutral gas observed at this epoch 
basically gone into stars in disks and irregulars.
 
In our model the residual gas left in galaxies is 
$\Omega_{\rm gas}=0.0002$.
This is compared with $\Omega_{\rm HI+HeI}=0.0003$ from an HI survey
of Rao \& Briggs (1993).  This residual gas still causes star
formation with a rate of $0.016 M_\odot h^3$ Mpc$^{-3}$ yr$^{-1}$
which is compared to $0.026{+0.014 \atop -0.010} M_\odot h$
Mpc$^{-3}$ yr$^{-1}$ obtained for the local sample 
by Gallego et al. (1995) from an H$\alpha$ survey.
The star formation rate in our model increases rapidly to
$z\sim 1$ (by a factor of 2-3). Such an increase is 
observed by Cowie, Hu \& Songaila (1995) and by Lilly et al. (1996), but
quantitatively, our increase between $z=0$ and 1 is about 1/2--1/3 that
reported by Lilly et al. (1996). 
 
%
%

\section{Discussion}
 
We have presented a simple, but quantitative model for the
history of galaxies. Most of the ingredients are not particularly
new, known for a while from studies of many authors,
although the interpretations may differ from author to author.
Here, we have assembled them into a single model in a
consistent way. The model is reasonably successful to reproduce many
different features observationally found. Many aspects are checked with
observations for $z\lsim 1$, and hence the model is well constrained.
On the other hand, the model is not well constrained for higher $z$, due
to the lack of observations other than the number count data.
 
An almost inevitable consequence of the present scenario is that
morphology of galaxies becomes manifest only in a
later epoch. A likely
control parameter is spheroid mass (Meisels \& Ostriker 1984),
or the virial temperature of
spheroids, besides environment. We might speculate that cooling is
not sufficient to make disks when the virial temperature is high,
leading to elliptical galaxies, and
disks only form when associated with lower mass spheroids. It is likely
that disk formation is hindered in cluster environments, since the
virial temperature is quite high in clusters. Elliptical galaxies
generally form from high $\sigma$ peaks of the density 
fluctuations, which makes the
two point correlation of elliptical galaxies automatically high.
In this scenario the cutoff of the elliptical LF 
in its faint end is obvious. Above $z\gsim 2$ disk galaxies
would be observed as ``naked bulges''; fuzzy cool cloud may
eventually be assembled around the bulges as time goes on, and
finally show up the full disk structure by $z\sim 1$. This
implies that the size of disk galaxies was substantially
smaller at $z\ge 1$.
 
In our model mergers do not play a major role, although the
presence of the modest amount of mergers is not precluded. There is
much evidence that the merger rate  cannot be as high as needed 
to explain the sharp increase of the $B$ count data. 
The disk of our galaxies cannot
acquire more than 4\% of its mass in the past 5 Gyr, otherwise the disk
scale height and Toomre $Q$ parameter would exceed the observed values
(T\'oth \& Ostriker 1992). Much evidence
from colour and emission properties of normal galaxies suggests that
the star formation has persisted over the Hubble time without much
change (Kennicutt 1983; Gallagher, Hunter \& Tutukov 1984). 
This supports the view that galaxy properties have not undergone 
much change, as expected in the ``major mergers'', 
at least after $z\sim 1$.
(``Minor mergers'', e.g., accretion of small galaxies onto the disks
might have taken place though; see below.)
A high merger rate (Guiderdoni \& Rocca-Volmerange 1991; Broadhurst
et al. 1992) leads to a trend of decreasing redshift for fainter 
magnitude for redder colour bands, which however is not 
supported by the recent deep redshift
survey ($I\lsim 24$) from the Keck telescope (Koo 1996). 
More directly, the Mg II absorption-line selected sample
indicates that properties of giant galaxies change very little
after $z\sim 1$, indicating no evidence for bulk merging after 
this redshift.
No data, however, can constrain the importance of mergers 
for $z\gsim 1$. Where
only the count data are available, an increase in  numbers and a
decrease in luminosity compensate each other and modest merging
does not modify the predictions. Therefore, the success of the
present model does not necessarily mean the absence of merging for
$z>1$. This question is still not settled.
 
The epoch of the spheroidal formation is also a long standing problem.
In our model, we simply assumed that the spheroid formation took
place at a high redshift. An underlying idea for the early formation
of spheroids is that stars must have formed within the collapse
time and hence the first collapse of massive objects may be a
suitable place for spheroid formation. The rise of the quasar population
at $z=4$ to 3 may also imply that spheroids formed at least earlier than
quasars get active (Rees 1995).
If, indeed, most of spheroids formed below
$z<5$, the light from the star formation epoch should be visible even in
a $K=18$ sample. Therefore, it will be particularly interesting to
find spectroscopic
features of those galaxies in the $K$ selected sample (Songaila 
et al. 1994)
which have been  missed in their spectroscopic follow-up.
Perhaps, the technique similar to that
developed by Steidel et al. (1996) can  most efficiently be applied
to such a sample.  If the result excludes the possibility of these
galaxies having $z=3-5$, the remaining possibilities are either (i) they
formed at very high $z$, or (ii) the formation redshift of spheroids
is not so high, and they formed at around the same time as galactic
disks, i.e., $z\lsim 3$. We note that the baryons in
spheroids amount to 2/3 (somewhat less in case of late spheroid
formation) of total
baryons frozen into stars. Then, if (ii) is true, the 
integrated light from $z=1-3$ must
be more than $\approx 5$ times higher than is predicted here, 
since star formation must have been stopped 
at some redshift above $z=1$. 
Such activity can easily be detected, e.g., especially with use
of narrow band filters selecting for H$\alpha$ emission.
Of course, this modifies the basic assumptions taken here and
hence most of the predictions.
 
The evolution of spheroids is reasonably well
understood, although quantitative details depend on models.
The poorly-understood part is the UV spectrum. There are a number
of speculations about the origin of the UV component observed in
elliptical galaxies today, but there seems to be no consensus.
In this paper we have passed over this problem.
What we have discussed in this paper is the prediction that 
is not sensitive to the
presence or absence of the UV component of spheroids.  One
specific case where such UV is important is  seen in the
high $z$ tails of the redshift distributions of optically selected
samples, the most notable example being ``Lyman continuum break
galaxies'' having $z=3-3.5$ in the $R=22-23.5$ mag sample, as discovered
by Steidel et al. (1996). In this case the observability of these
galaxies is completely controlled by the UV light emitted from 
spheroids. In the present model, we predict that the tail of the 
redshift distribution of $R=22-23.5$ sample does not extend 
beyond $z\simeq 2.5$.
The initial burst of spheroids does not give enough a number of
Lyman break galaxies, unless the formation epoch is lower than $z=4$.
If, however, some star formation activities persist, e.g., those
with recycled gas, it is easy to give a sufficient number of such
galaxies, without modifying any other predictions discussed in this
paper.
 
The evolution of spiral galaxies is less well understood. There
are, however, a number of constraints from observations. Among them,
the presence and strength of high redshift tail in the redshift 
distribution of a $B$ selected sample serve as an interesting 
indicator for the amount of evolution of giant spiral galaxies. 
While it is generally
agreed that the tail is not very large, it is still not settled
observationally to what degree such a tail is present (see Cowie et al.
1996; Koo 1996).
A very strong constraint
comes from the Mg II sample, which precludes any appreciable
luminosity evolution
of spiral galaxies in the relevant redshift interval. On the other hand,
the CFRS survey (Lilly et al. 1995) does see evolution of the 
characteristic luminosity $L^*$ for blue, presumably spiral,
galaxies by one magnitude (such evolution is not claimed 
in Ellis' et al. 1996 work, though). 
In our model these two constraints are somehow
reconciled in the way that evolution is faster between $z=0$ 
to $z=0.3-0.5$, but very slow above this redshift. 
This is because gas infall has almost
ceased by $z=0.5$, and the galaxies started to show more evolution after
this redshift. Such evolution constraints, if taken literally, indicate
that a spiral galaxy is not a closed system, but supplied with fresh gas
down to $z=1$ or less. 
 
One of the main points of the present paper is to estimate the age of
irregular galaxies and introduce them as being newly formed 
at low redshift.
They are the agent that is responsible for the steep slope of the number
count in the blue bands. These galaxies are quite gaseous, and undergo
very active star formation, which may be observed as
star bursts. A similar idea has already been advocated by
a number of authors, in particular by Cowie et al. (1991)
and Babul \& Rees (1992).
The difference is in the point that we have not
introduced any population unobservable today and
that we assume continuous star formation that is proportional
to the gas abundance rather than the ``burst''.
The dwarf population, which is less luminous
than ordinary irregular galaxies, plays no role in the results
presented in this paper. There may be more irregular galaxies formed
at higher $z$, than are considered in this paper, but they should 
follow the same track as the disk component of spiral galaxies, and may
constitute a low-luminosity tail of the LF of spiral galaxies, which
is not important in galaxy number counts. The detailed shape
of the subluminous part of the luminosity function is not important,
unless the LF is very steep.  

There is much indication that
correlation of galaxies seen in blue bands is weak (Efstathiou et al.
1991; Couch, Jurcevic \& Boyle 1993; Infante \& Pritchet 1995), 
the clustering being
at most comparable to that of late type galaxies. This is easy to
understand: the $B$ sample at faint magnitudes is dominated by
spiral galaxies and irregular galaxies which form from the 
peaks of low-$\sigma$ fluctuations and are thus least biased.
 
The local LF is clearly one of the most important sources of 
uncertainties, and it often hinders us from making unambiguous
predictions from our model.  The uncertainty in the local LF
is not only in the faint end behaviour, but also in its normalization.
An upward shift of the normalization by 0.15 dex would bring a number
of predictions of the galaxy counts in much better agreement 
with observations, and also it would decrease substantially 
the amount of evolution required by the high 
redshift data. Our normalization of the
local LF, however, is on the high side of
what is allowed by current surveys, 
so we cannot justify a further increase by 0.15 dex.
More important is the
problem of the faint end ($M_{\rm B} \sim -16$ mag) 
of the LF. If the total LF has a flat slope
($\alpha\sim-1$) as in, e.g., Loveday et al. (1992) and 
Lin et al. (1996), that is compelling evidence for
galaxy number evolution. Luminosity evolution does not have enough power
to explain the abundance of irregular galaxies required in our
analysis. This view is consistent with the LF at various redshifts
derived by Ellis et al. (1996): the luminosity evolution does not
have power to lift the faint end of the LF. 
On the other hand, if the local luminosity
function shows an increase towards the faint end (Metcalfe et al. 1991;
Marzke et al. 1994; SubbaRao et al. 1996), we do not particularly 
need number evolution of galaxies, but luminosity evolution is
sufficient to explain the observations. 
It is also possible that irregular
galaxies have evolved into dwarf spheroidals near giant galaxies
or in a galaxy rich environment. In this case the
rise of the LF may take place $\sim$1 mag fainter than the 
typical luminosity of irregular galaxies.

We have argued that $K$ band observations provide a good tool to study
cosmology. This is because spheroids always contribute significantly
to the $K$ band counts
and the evolution of spheroids is reasonably well
known. We have concluded that $\Omega=1$ is disfavoured by the count
data. This conclusion remains true, as long as we keep our basic
assumptions. If, however, mergers play an important role at high $z$,
say $z\simeq 2$, the $\Omega=1$ model could be allowed. The $z$
distribution for $\Omega=1$ cosmology has a component substantially
extended to higher $z$ than in $\Omega=0.1$ model.
The median redshift of the spheroids in a $K\simeq 22$ sample is about 3
(for an $\Omega=0.1$ model), 
and an observation with $U$ and $G$ bands
(as in Steidel et al. 1996) searching for Lyman break
galaxies just hits the middle of the $z$-distribution at this magnitude.
The possibility that not all spheroids are
in place by $z\sim3$ may also modify the conclusion about cosmology,
and it may bring a confusion as to the interpretation  of the
cosmological models.
 
%
%

\acknowledgments
 
We should like to thank  Nobuo Arimoto, Richard Ellis, Roger Davies, 
Tadayuki Kodama, Simon Lilly, Richard MacMahon and 
Jim Peebles for useful discussions, and Jarle Brinchman, 
Richard Ellis and Ofer Lahav for reading manuscripts
and many comments improving the manuscript. One of us (MF) thanks       
the hospitality of the Institute of Astronomy of the University of 
Cambridge, where this paper is made into the final form. He also
wishes to acknowledge generous support from the Fuji Xerox Corporation
at Princeton.  
This research was supported in part by 
the Grants-in-Aid by the Ministry of Education, 
Science, Sports and Culture of Japan (07CE2002) 
to RESCEU (Research Center for the Early Universe).
 
%
%

\end{document}